\documentclass[manuscript]{aastex}

\addtocounter{footnote}{1}

\slugcomment{Accepted for publication in The Astrophysical Journal}

\shorttitle{Solar origin of the solar energetic particle event on 2013 April 11}
\shortauthors{Lario et al.}

\begin{document}


\title{The Solar  Energetic Particle Event on 2013 April 11:\\
     An Investigation of its Solar Origin and  Longitudinal Spread}


\author{D. Lario\altaffilmark{1} and N.E. Raouafi\altaffilmark{1}}
\affil{The Johns Hopkins University, Applied Physics Laboratory,
    Laurel, MD 20723}

\author{R.-Y. Kwon\altaffilmark{2} and J. Zhang\altaffilmark{2}}
\affil{School of Physics, Astronomy and Computational Sciences, George Mason University,
    Fairfax, VA 22030}

\author{R. G\'omez-Herrero\altaffilmark{3}}
\affil{Space Research Group, Physics and Mathematics Department, University of Alcal\'a, Alcal\'a de Henares, E-28871 Spain}

\author{N. Dresing\altaffilmark{4}}
\affil{Institute of Experimental and Applied Physics, Christian-Albrechts University of Kiel, Kiel, Germany}

\and

\author{P. Riley\altaffilmark{5}}
\affil{Predictive Science, 9990 Mesa Rim Road, Suite 170, San Diego, CA 92121}



\altaffiltext{1}{The Johns Hopkins University, Applied Physics Laboratory,
    Laurel, MD 20723}
\altaffiltext{2}{School of Physics, Astronomy and Computational Sciences, George Mason University, 4400 University Dr., MSN 6A2, Fairfax, VA 22030}
\altaffiltext{3}{Physics and Mathematics Department, University of Alcal\'a, Alcal\'a de Henares, E-28871 Spain}
\altaffiltext{4}{Institute of Experimental and Applied Physics, Christian-Albrechts University of Kiel, Kiel, Germany}
\altaffiltext{5}{Predictive Science, 9990 Mesa Rim Road, Suite 170, San Diego, CA 92121}


\begin{abstract}
We investigate the solar phenomena associated with the origin of the solar energetic particle (SEP) event observed on 2013 April 11 by a number of spacecraft 
distributed in the inner heliosphere over a broad range of heliolongitudes. 
We use Extreme UltraViolet (EUV) and white-light coronagraph observations from the {\it Solar Dynamics Observatory} (SDO),
the {\it SOlar and Heliospheric Observatory} (SOHO) and the twin {\it Solar  TErrestrial RElations Observatory} spacecraft (STEREO-A and STEREO-B)
 to determine the angular extent of the EUV wave 
and coronal mass ejection (CME) associated with the origin of the SEP event.
We compare the estimated release time of SEPs observed at each spacecraft with the arrival time of the structures associated
with the CME at the footpoints of the field lines
connecting each spacecraft with the Sun.
Whereas the arrival of the EUV wave and CME-driven shock at the footpoint of STEREO-B is consistent, within   uncertainties,
with the release time of the particles observed by this spacecraft, the EUV wave never reached the footpoint of the field lines connecting near-Earth observers
with the Sun, even though an intense SEP event was observed there.
We show that the west flank of the CME-driven shock propagating at high altitudes above the solar surface
was most likely the source of the particles observed near Earth, but it did not leave any EUV trace on the solar disk. 
We conclude that the angular extent of the EUV wave on the solar surface did not agree with the longitudinal extent of the SEP event in the heliosphere. 
Hence  EUV waves cannot be used reliably as a proxy for the solar phenomena that accelerates and injects energetic particles over broad ranges of longitudes.
\end{abstract}





\section{Introduction}

The fleet of spacecraft distributed throughout the inner heliosphere during  solar
cycle 24 offers us the unique opportunity to study solar energetic particle (SEP) events from
multiple vantage points \citep[e.g.,][]{dresing12, dresing14, rouillard12, lario13, richardson14}.
The combination of in-situ and remote-sensing observations from a number of detectors
onboard these spacecraft such as those on the SOlar and Heliospheric Observatory \citep[SOHO;][]{domingo95}
and on the Solar TErrestrial RElations Observatory \citep[STEREO;][]{kaiser08} has proven to be a valuable tool to identify
the solar processes associated with the origin of the SEP events as observed from different heliolongitudes
\citep[e.g.,][]{rouillard12, park13, prise14}.
However, the connection between SEP event in-situ measurements
and remote-sensing observations of the solar phenomena typically associated with the origin of the SEP events, such as 
large-scale coronal propagating fronts, has not been always conclusive.

For example, 
\citet{bothmer97} and \citet{Posner97} analyzed 0.25-0.7 MeV electron intensities
observed by SOHO during the SEP event on 1997 April 7.
This event was associated with a C6.8 X-ray flare at S30E19. A  wave front observed by the Extreme Ultraviolet Imaging Telescope 
\citep[EIT;][]{delaboudiniere95} onboard SOHO
(usually referred to as an EIT wave) was observed to extend from this flare site toward the nominal 
magnetic connection footpoint of this spacecraft. However,
the onset of the electron intensities at SOHO occurred earlier than the EIT wave reached the western solar hemisphere
where, in principle, SOHO was magnetically connected.
The conclusion drawn by these authors was that EIT disturbances in the low corona 
were unlikely sources of the SEP events observed by SOHO
\citep{bothmer97, Posner97}.

On the other hand, \citet{torsti99} analyzed the $\sim$11-100 MeV proton event observed by SOHO 
on 1997 September 24 associated with a solar flare at S31E19
(i.e.  $\sim$88$^{\circ}$ eastward from the SOHO's nominal magnetic footpoint).
\citet{torsti99} considered the EIT wave as a signature of the expansion
of the solar phenomena associated with the origin of the 
 $^{>}_{\sim}$10 MeV proton intensity enhancement observed at SOHO.
As stated by the authors, the EIT wave was considered only the ``ground track'' or ``moving skirt'' of the expanding
coronal shock but not the agent responsible for the particle acceleration.
Whereas the EIT wave emission usually occurs within $\sim$0.2 R$_{\odot}$ of the solar surface,
\citet{torsti99} suggested that the particles were accelerated at higher coronal altitudes,
probably at $\sim$0.5-1 R$_{\odot}$ from the solar surface,
where they were  injected onto open field lines  extending into interplanetary space.

\citet{krucker99} investigated the origin of impulsive electron events at energies $>$25 keV and attributed
electron events with delayed onsets, mostly from nominal poorly-connected regions, to
the time for coronal waves to propagate from the site of the parent active region to well-connected
longitudes.
However, it was indicated that the observed part of the wave fronts at low coronal altitudes were moving too
slowly to connect to the footpoint of the field line connecting with the spacecraft,
but at higher altitudes ($\sim$1.5 R$_{\odot}$) the speed of the wave fronts could be fast enough to explain the release
of the electrons from  higher regions in the solar corona.

More recently, \citet{miteva14} analyzed the link between EIT waves and particle events in 179 SEP events
observed by SOHO during solar cycle 23.
Whereas most of the SEP events (87$\%$) were accompanied by EIT waves,
for those events generated from eastern longitudes,
the extrapolated arrival time of the EIT waves to the Earth's magnetic connection point was consistent with the 
onset of 25-41 MeV protons.
However, on a number of events, the first 0.25-0.7 MeV and 0.67-3 MeV
electrons were detected too early, ruling out the possibility that the expansion of the EIT wave marked the release of electrons for the longitudes
well connected to SOHO.

With the advent of multi-spacecraft in-situ and remote-sensing observations provided by   STEREO and by near-Earth spacecraft,
we have the possibility to study the longitudinal extent of individual SEP events and their link
with the angular extension of the solar phenomena typically associated with the acceleration of SEPs in the solar corona.
For example,
\citet{rouillard12} studied the SEP event observed on 2011 March 21 using a combination of data from SOHO at the first Earth-Sun Lagrangian point (L1) and
from STEREO-A and STEREO-B (separated by 88$^{\circ}$ and 95$^{\circ}$, west and east of Earth, respectively).
\citet{rouillard12} determined that there was an association
between the extent of the SEP event in the heliosphere and the longitudinal extent of the perturbed corona.
By combining Extreme UltraViolet 
(EUV) and white-light images of the lower corona, \citet{rouillard12} determined that during
the initial CME lateral expansion, an EUV disturbance propagated parallel to the solar surface
tracking the lateral expansion of the ejecta. The arrival time of the EUV disturbance$^{1}$  at the longitudes
of the footpoints of the nominal magnetic field lines connecting SOHO and STEREO-A with the
Sun coincided (within 12 minutes) with their estimate of the  release time of the SEPs
observed by each spacecraft. The EUV wave never reached the estimated magnetic footpoint of
STEREO-B and hence \citet{rouillard12} concluded that SEPs were not able to reach STEREO-B.
However, it is worth pointing out that \citet{richardson14} noticed a small increase in the $\sim$25 MeV proton intensities at STEREO-B associated with the event on 2011 March 21.

\footnotetext{The coronal bright fronts observed in extreme ultraviolet images as moving EUV disturbances have been proven to be wavelike phenomena 
\citep[e.g.,][]{long08, patsourakos09, olmedo12}. These bright fronts were originally termed EIT waves because of their detection with SOHO/EIT \citep[e.g.,][]{thompson98}, but when using STEREO
observations have been usually referred to as EUV waves. Henceforth we will use the term EUV wave.}

Recently, \citet{park13} studied the source regions of 12 SEP events by combining 195 {\AA} observations from
the Extreme Ultraviolet Imager \citep[EUVI;][]{wuelser04} of the Sun Earth Connection Coronal and Heliospheric Investigation 
\citep[SECCHI;][]{howard08} onboard the STEREO and 193 {\AA} images from the Atmospheric Imaging Assembly \citep[AIA;][]{lemen12} onboard
the Solar Dynamics Observatory \citep[SDO;][]{pesnell12}. 
The arrival time of the EUV waves at the magnetic footpoints of the spacecraft were compared to the SEP onset times of 175-315 keV and 335-375 keV electron
intensity enhancements and 1.9-3.06 MeV and 1.8-3.6 MeV proton intensity enhancements.
In order to estimate the arrival time of the EUV waves at the connecting magnetic footpoints,  \citet{park13} used 
running ratio images from STEREO/EUVI and SDO/AIA and visually determined when the waves arrived at the spacecraft footpoints or alternatively
assumed a linear extrapolation of the EUV wavefronts in the running  ratio space-time images to the foopoint sites. 
Note that this alternative method ascribes an
arrival time of the EUV wave to a given footpoint although the actual arrival of the wave at the footpoint is not always observed.
\citet{park13} pointed out that in some events, although SEPs  were recorded by some spacecraft, the EUV wave did not reach the respective connection footpoints.
In spite  of the poor correlations found between the EUV arrival times and the SEP onset times (see their Figures 11-13), 
\citet{park13} concluded that SEP onset times were significantly associated with the EUV arrival times,
suggesting that EUV waves trace the release sites of SEPs.
However, in contrast to protons, the delay time between electron onset and the EUV wave reaching the connecting footpoint was
found to be independent of the distance from the flare site.

More recently, 
\citet{prise14} studied the evolution of the CME and EUV wave associated with  the origin of the SEP event on 2011 November 3
observed by both STEREOs and near-Earth spacecraft.
They found that the initial lateral expansion of the CME low in the corona closely tracked
the propagation of the EUV wave.
However, the EUV wave was not observed to reach the footpoints of any of the spacecraft at their respective energetic particle release time.
Coronagraph observations showed that the white-light CME, higher up in the corona, reached the field lines connecting the two STEREO spacecraft at a 
time consistent with the release of particles.
In contrast to the study by \citet{rouillard12}, \citet{prise14} concluded that the propagation of the EUV wave 
over the solar surface could not be used as a proxy for  the 
expansion of the CME at high altitudes since it was not observed to reach
the magnetic footpoints of some of the spacecraft observing SEPs.
The conclusion drawn by \citet{prise14} was that 
the factor that determines the longitudinal expansion of the SEP events is
the CME
at high altitudes but not the EUV wave in the low corona.

In this article we study the first large Fe-rich SEP event of solar cycle 24 that took place on 2013 April 11 \citep{cohen14}.
The site of the solar flare associated with the origin
of the SEP event (N09E12)  allows us to study in detail the associated EUV wave with the
high cadence and resolution data provided by SDO/AIA.
We compare the estimated release time of the first observed particles with the arrival of the EUV wave and white-light coronal shock at the magnetic connection footpoints
of each spacecraft that observed the event.
In Section 2 we present the observations associated with this event.
In Section 3 we discuss the reliability of using EUV waves as a proxy for the longitudinal extent of
the SEP events and the observed differences in the energetic electron and proton release times.
Finally, in Section 4 we summarize the main results of the present study.
 
\section{Observations} 
 
\subsection{Spacecraft locations and magnetic connection footpoints}
 
 Figure~1 shows the longitudinal distribution of spacecraft located at heliocentric distance $R$$\sim$1 AU on 2013 April 11 as seen from the north ecliptic pole.
 The red, blue and black dots indicate the locations of STEREO-A (STA), STEREO-B (STB) and observers near the Sun-Earth Lagrangian point
 L1, respectively.
 $R$ indicates the heliocentric radial distance of the spacecraft, and $Long$ their heliographic inertial longitude.
 This distribution of spacecraft allows us to determine whether this SEP event spread over a wide range of heliolongitudes. 
 Table~1 lists the locations of the spacecraft and the footpoints of the field lines connecting each spacecraft with the Sun estimated
 using different methods as described below.
 Columns 2, 3, 4 and 5 of Table~1 list the heliocentric radial distance ($R$), the heliographic inertial longitude ($Long$), the Carrington Longitude ($CL$)
 and the heliographic inertial latitude ($Lat$) of STEREO-B, Earth, and STEREO-A,
 respectively. 
We have added in Figure~1, 
the longitude of the associated parent flare indicated by the purple line (E12 as seen from Earth, or 113$^{\circ}$ in terms of heliographic inertial longitude) and
 nominal Parker spiral field lines connecting each spacecraft with the Sun considering the solar wind speed $V_{sw}$ measured at the onset of the SEP event
 (as listed in Column 6 of Table~1).
 The coordinates of the magnetic connection footpoints for each spacecraft along these nominal field lines are listed in columns 7 and 8 of Table~1. 
 The longitudinal distance between these footpoints and the flare site is  
 $\Delta\psi_{STA}$$\sim$171$^{\circ}$, $\Delta\psi_{STB}$$\sim$58$^{\circ}$, and $\Delta\psi_{Earth}$$\sim$-73$^{\circ}$
 (where negative values indicate that the flare is eastward of the nominal footpoint).
 The  distance spacecraft-Sun along these nominal field lines is  $L_{STA}$$\sim$1.04 AU,  $L_{STB}$$\sim$1.25 AU,  and $L_{Earth}$$\sim$1.16 AU
 for STEREO-A (STA), STEREO-B (STB) and Earth (or L1 observers), respectively.
 
 The location of the magnetic connection footpoints can also be estimated using 
 the results of the ``Magnetohydrodynamics outside A Sphere model"
 \citep[i.e. the MAS model;][]{riley12}.
Such model solves the time-dependent, resistive MHD equations in spherical coordinates, and describes
 the large-scale behavior of the solar corona and inner heliosphere from the solar surface to 30 R$_{\odot}$.
It uses
photospheric magnetic field synoptic maps built up from a
sequence of observations from the Helioseismic and Magnetic Imager \citep[HMI;][]{scherrer12} onboard SDO  
and determines  the large-scale magnetic field structure of the corona.
The details of the models can be found elsewhere
\citep[e.g.][and references therein]{riley12}.
Figure~2 shows the magnetic field configuration of the corona as seen from Earth at 06:04 UT on
11 April 2013. This time is immediately before the occurrence of the parent solar flare and CME associated with the origin of the SEP event (see Section 2.3). 
The spherical surface, at 1 R$_{\odot}$ in Figure~2, shows the location and polarity of the coronal holes
(red for outward magnetic field and blue for inward). A selection of closed (green) and open
(gold) field lines is also shown. 
The site of the solar flare associated with the origin of the SEP event (N09E12) is indicated by the orange cross in Figure~2.

In order to estimate the magnetic connection footpoints of each spacecraft using the results of the MAS model,
we use the solar wind speed
measured by each spacecraft (column~6 of Table~1) to ballistically project
the plasma measured in situ  to a distance
30 R$_{\odot}$. This field line is then mapped
back to the solar surface according to the magnetic
configuration given by the MHD simulation.
Figure 3 shows the coronal holes for Carrington Rotation 2135
(from day 80.4971 to day 107.7802 of 2013)
computed from the MHD solution and color-coded
according to the observed underlying photospheric
field (red for outward and blue for inward).  
The units in the horizontal axis are degrees of the Carrington Rotation 2135 that can be transformed into heliographic inertial longitude
by just adding 40$^{\circ}$. The Earth's trajectory is superimposed (at
latitude -6$^{\circ}$) together with the mapped source
regions of the plasma according to the solar wind
speed measured at ACE. The thin green lines
establish the connection between the ballistic
projection at 30 R$_{\odot}$ and the connecting point at 1 R$_{\odot}$
through the field lines computed by the MHD
model. Columns 9 and 10 of Table~1 shows the estimated coordinates of the  magnetic connection footpoints of the three spacecraft using this method
and indicated by diamonds in Figure~3 (blue for STEREO-B, black for Earth and red for STEREO-A).
 
 Additionally,
  we have also considered  the  method used by \citet{park13}
to estimate the location of the magnetic connection footpoints.
A ballistic projection of the interplanetary magnetic field through a Parker spiral using the instantaneous measured solar wind speed (column 6 in Table~1) up to a distance of 2.5 R$_{\odot}$ 
is then followed by the open field lines resulting from
the potential field source surface (PFSS) model applied to the HMI magnetogram obtained at 06:04 UT on 2013 April 11.
Columns 11 and 12 of Table~1 show the estimated coordinates of these footpoints using the PFSS method.

Note that  all these three methods (i.e., Parker spiral projection, MAS and PFSS) provide the best estimation (in terms of the models and means currently available) 
of the magnetic connecting footpoint locations prior to the eruption that generates the SEP event.
It is well understood that as soon as the parent eruption occurs, the magnetic field in the low corona may drastically change.
The location of the magnetic footpoints on the solar surface using these three methods gives us an estimation of
the longitudinal distance between the site of the active region generating the event and the root of the field line where, in principle, the 
first arriving energetic particles should be released to be observed by the spacecraft. 
For this specific event, the period prior to the solar eruption generating the SEP event on 2013 April 11 was exceptionally quiet:
no $>$600 km s$^{-1}$ CMEs were observed for a period of 10 days before the onset of the SEP event and 
no X-ray flares of class above M were observed for a period of 6 days prior to the SEP event.
The last $>$600 km s$^{-1}$ CME prior to the CME associated with the event on 2013 April 11 was observed at 02:36 UT on 2013 April 1 according to the CME LASCO catalog
posted on cdaw.gsfc.nasa.gov/CME$\_$list.
The last X-ray flare of class above M prior to the onset of the SEP events occurred at 17:34 UT on 2013 April 6 (www.swpc.noaa.gov).

The main differences among the magnetic footpoint coordinates listed in Table~1 for  the three spacecraft are (1) 
the latitude of the footpoints obtained using either the Parker spiral projection or  the MAS and the PFSS methods, 
and (2) the different longitudes assigned to STEREO-B's footpoint by the three methods.
 Whereas
ballistic projection through a Parker spiral does not consider
latitudinal excursions with respect to the spacecraft latitude,  both MAS and PFSS allow for latitudinal excursions along 
the open field lines close to the corona provided by the models.
The longitudes of the footpoints associated with STEREO-B  differ by as much as 20$^{\circ}$ among the three different methods.
The presence of a small coronal hole with open field lines expanding over a wide range of longitudes
(at $CL$=0$^{\circ}$ in Figure~3) is the main reason for this discrepancy.

\subsection{Solar energetic particle observations}

Figure~4 shows a collection of energetic particle intensities measured by
(from left to right) STEREO-B, near-Earth observers,
and STEREO-A
during the event on 2013 April 11 (day of year 101).
From top to bottom we show one minute averages of $>$13 MeV proton intensities in several differential energy channels
(Figure~4a), near-relativistic and relativistic electron intensities (Figure~4b), and hourly averages of $\sim$18 MeV/n
Fe and O intensities (Figure~4c).
Proton intensities (Figure~4a)  were measured by the High-Energy Telescope  \citep[HET;][]{rosenvinge08} onboard STEREO
and by the Energetic Relativistic Nuclei and Electron instrument \citep[ERNE;][]{torsti95} onboard SOHO.
Near-relativistic electron intensities (red traces in Figure~4b) at both STEREO-A and STEREO-B were measured by the 
Solar Electron and Proton Telescope  \citep[SEPT;][]{muller08} whereas at the Advanced Composition Explorer (ACE) were measured by the 
Deflected Electron (DE) system of the Electron, Proton, and Alpha Monitor \citep[EPAM;][]{gold98}.
Relativistic electron intensities (blue trace in Figure~4b) at the two STEREO spacecraft were measured by the HET experiment
and near Earth by the Electron Proton Helium instrument \citep[EPHIN;][]{muller95} onboard SOHO.
Finally, the Fe and O intensities (Figure~4c) at the two STEREOs were measured by the Low Energy Telescope \citep[LET;][]{mewaldt08}
and at ACE by the Solar Isotope Spectrometer  \citep[SIS;][]{stone98}.

STEREO-B and L1 observers detected a substantial SEP event
with a rapid intensity increase followed by a gradual decay (in fact this SEP event was also classified as a NOAA SEP event
showing a $>$10 MeV proton peak intensity of 114 protons (cm$^{2}$ s sr)$^{-1}$;
c.f. www.swpc.noaa.gov/ftpdir/indices/SPE.txt).
This event also showed high Fe/O ratio at both STEREO-B and L1 observers  \citep[e.g.,][]{cohen14}.
At the time of the maximum peak intensity the  $\sim$18 MeV/n Fe/O ratio was
$\sim$0.60 at STEREO-B and $\sim$0.34 near-Earth.
Details of Fe/O ratios at different energies averaged over the duration of the SEP event as observed by STEREO-B and ACE can be found in \citet{cohen14}.
STEREO-A proton intensities did not show a significant increase  beyond the 
1, 2 or 3 discrete count rates, whereas  Fe or O intensities at this spacecraft did not show any increase.
However, near-relativistic electron intensities at STEREO-A showed a gradual increase of nearly one order of magnitude
with respect to the pre-event intensity.

The SEPT instrument onboard the  three-axis stabilized STEREO spacecraft allows us to estimate 
energetic particle pitch-angle distributions during the SEP events \citep[e.g.,][]{dresing14}.
SEPT consists of four identical telescopes mounted to cover four viewing directions
which are to the north, to the south, along the nominal Parker spiral to the Sun and 
along the nominal Parker spiral  away from the Sun \citep{muller08}.
The coverage of pitch-angles by these four telescopes depends on the orientation
of the magnetic field.
Similarly, the multiple angular sectors used by the 3D Plasma and Energetic Particle instrument (3DP) on board the spin-stabilized  WIND spacecraft  \citep{Lin95}
as well as the eight angular sectors of the Low Energy Foil Spectrometer (LEFS60) of the EPAM instrument onboard the spin-stabilized ACE spacecraft  \citep{gold98}
allow us to determine pitch-angle distributions of near-relativistic electrons near Earth.
In particular we use (i) $\sim$25-400 keV electron intensities measured by the Silicon Semiconductor Telescopes (SST) of WIND/3DP stored into eight  pitch-angle bins
and downloaded from 
sprg.ssl.berkeley.edu/wind3dp/, and (ii) $\sim$53-175 keV electron intensities measured by the ACE/EPAM/LEFS60 telescope sectored into eight viewing directions.

Figure~5 shows the near-relativistic electron anisotropy information
at the onset of the event as seen by (from left to right) STEREO-B, WIND, ACE and STEREO-A.
The top colored panels show electron pitch-angle distributions
color coded
according to the measured intensity
(indicated by the top color bar) and divided
into 4 pitch-angle bins. 
The second panel shows the orientation in pitch-angle space of the center point of the field of view of each telescope 
in the case of STEREO/SEPT, of each sector in the case of ACE/EPAM/LEFS60, and of each of the 8 pitch-angle bins provided in the 
WIND/3DP/SST data files.
The third panel shows the  intensity measured in each one of the four telescopes in the case of STEREO/SEPT,
in each sector of ACE/EPAM/LEFS60, 
and in each one of the 8 pitch-angle bins in the case of WIND/3DP/SST.
The fourth panel shows the first-order anisotropy coefficient
defined as 
$A=3\int_{-1}^{+1} I(\mu)\mu d\mu/\int_{-1}^{+1} I(\mu) d\mu$
where $I(\mu)$ is the pitch-angle dependent intensity
measured in a given viewing direction and $\mu$ is the average pitch-angle cosine
for that direction.
Details of how this anisotropy coefficient is computed can be found in \citet[and references therein]{dresing14}.
The sign of $A$ is given in terms of the magnetic field polarity. 
For that reason we show in the bottom of Figure~5 the 
magnetic field magnitude and magnetic field angular directions at the onset of the event as measured, from left to right,
by STEREO-B, WIND, ACE and STEREO-A (RTN coordinates are used for ACE and the two  STEREOs, whereas GSE coordinates are used for WIND). 
According to the magnetic field polarity registered at the onset of the event,  
positive values of $A$ in STEREO-B
indicate particles flowing in the anti-sunward direction, whereas negative values of $A$ in STEREO-A indicate particles flowing
in the anti-sunward direction.
The flow of particles at the onset of the electron event at both ACE and WIND was mostly anti-sunward, followed by a period where $A$ changed its sign, 
and a subsequent recovering of antisunward flow for the rest of the event.
Owing to the fact that this change of sign in $A$ was observed first at WIND (closer to the Sun than ACE)
and that the delay of field signatures between WIND and ACE (such as  changes in field orientation seen first by WIND and later by ACE) coincides with the solar wind convection time,
we suggest that the change of sign in $A$ was the result of a spatial structure moving over L1.

 The onset of the event at STEREO-B was highly anisotropic ($A$$>$2), relatively anisotropic at WIND and ACE ($A$$^{<}_\sim$1), 
 whereas the gradual increase at STEREO-A was mostly isotropic ($A$$^{<}_{\sim}$0.2).
In general,
the onset of SEP events generated from active regions magnetically well-connected to the observer
tend to show large anisotropies because of the prompt injection of particles into the magnetic field line
connected to the observer,  whereas the onset of the events generated from poorly connected longitudes,
usually characterized by slow gradual increases, show weak anisotropies
\citep[e.g.,][]{miteva14}.
We note here the  high anisotropic character of the event onset at STEREO-B   even when the longitudinal distance 
between the parent flare site and the nominal footpoint was more than $\sim$55$^{\circ}$ \citep{dresing14}.

The left panels of Figure~6 show in detail the onset of the SEP event at STEREO-B (Figure~6a) and SOHO (Figure~6b)
as observed in different proton energy channels of the HET experiment  on STEREO-B
and of the ERNE instrument on SOHO.
The black dots are one-minute resolution proton intensities and the red lines are 3-minute averages.
Note  the use of linear vertical scales to better visualize the onset of the event. The cleanliness of the instruments'
backgrounds allows us to see the event onset at the level of discrete counts.
The blue solid vertical lines in each panel indicate the onset of the event  identified as the first 3-minute interval where high-resolution
intensities do not return to the pre-event counts, and the intensities keep increasing.
For such kind of events with clear intensity increases,
alternative methods to identify onsets of SEP events 
\citep[e.g.,][and references therein]{torsti98, krucker99, huttunen05, vainio13, miteva14}
provide  results that do not differ significantly from those shown in Figure~6.
We should note that during this period, SOHO was rotated 180$^{\circ}$ with respect to its nominal pointing direction,
meaning that the fields of view of the particle instruments on SOHO (such as ERNE and EPHIN) 
were pointed perpendicular to the nominal Parker field direction
(cf. sohowww.nascom.nasa.gov/data/ancillary). 
Nominally, these two particle instruments point sunward along the nominal Parker field direction allowing detection of the first arriving particles
that tend to propagate along the field lines. Under this rotated configuration, it is possible  that,
in the case of highly anisotropic onsets, SOHO/ERNE missed the first 
protons that arrived at L1,
yielding  delayed onset times.

Figure 6c shows the results of the velocity dispersion analysis (VDA) applied to the event as
observed by STEREO-B. The VDA method consists 
in plotting the onset times at different energies versus $c/v=1/\beta$, where $v$ is the particle speed. 
The green diamonds indicate the onset times in the proton channels shown in Figure~6a.
The purple straight line in Figure~6c is a least-square fit to the onset times at different energies given by the expression
$t_{i}=A+D/v_{i}$ where $t_{i}$ is the onset time in the energy channel detecting particles of speed $v_{i}$ 
(considered to be the speed of particles with energy equal to the geometrical mean of the energy window of the channel),
$A$ is the release time of the particles at the Sun and $D$ the distance traveled by these particles.
The estimation of the release time by this VDA method assumes
(i) the onset of the SEP injection profile is assumed to be both
impulsive and energy independent, and (ii) the first arriving particles observed at the spacecraft are assumed to propagate scatter-free,
with pitch-angle cosines $\mu$$\sim$1, along a common travel distance $D$ from their coronal injection site to the observer. 
The intercept of the purple straight line in Figure~6c with the vertical axis gives an estimate of the release time of protons observed by STEREO-B 
at $A$=422$\pm$4 min 
(i.e. 07:10 UT $\pm$4 min adding $\sim$8 minutes to directly compare with the electromagnetic observations described below).
The long distance inferred by this method ($D$=2.38$\pm$0.45 AU)  may result from either a non-standard interplanetary magnetic field topology 
(longer than the expected length of $\sim$1.2 AU for a Parker spiral configuration)
or the fact that low-energy proton
onsets were considerably delayed with respect to high-energy proton onsets. The former can be ruled out
because of the lack of interplanetary transients observed prior to the event. The latter may be due to either a delayed
injection of low-energy protons and/or an energy-dependent transport process (e.g. more efficient scattering processes acting on low-energy protons).
Both processes would invalidate the use of the VDA method. 
Success and reliability of the VDA method typically used to estimate the release time of solar energetic particles have been recently
reviewed by several authors \citep[e.g.,][]{lintunen04, saiz05, kahler06, vainio13}. We refer the reader to these works (and references therein) for details on the reliability of this method.

We have added in Figure~6c the onset time of the electron intensities as seen in the sunward direction
of the highest electron energy channel of STEREO/SEPT (375-425 keV) and the lowest energy channel of STEREO/HET (0.7-1.4 MeV),
indicated by black dots in Figure~6c. We see that the electron onsets at STEREO-B are consistent with a simultaneous
release of both electrons and protons according to the VDA least-squares fit.
Just by time-shifting the onset time of the electron intensity enhancement along the nominal Parker spiral magnetic field line length $L_{STB}$ and considering the electron speed,
we estimate a release time for the relativistic (0.7-1.4 MeV) electrons of  07:17 UT $\pm$2 min,  and for the near-relativistic electrons (375-425 keV) of
07:16 UT $\pm$2 min. The same time-shifted analysis applied to the 60-100 MeV proton onset gives a release time of 07:26 UT $\pm$3 min
($\sim$8 minutes already added to compare with electromagnetic observations described below).

Figure~6d shows the onset times in the proton channels of SOHO/ERNE identified in Figure~6b as a function of $c/v=1/\beta$
(green diamonds).
It should be kept in mind that the incidental field of view of the instrument may prevent,
depending on the magnetic field orientation and the anisotropic character of the SEP onset,  the observation of the first arriving particles nominally aligned 
with the magnetic field.
The small  anisotropy at the onset of the SEP event at L1 (Figure~5) 
suggests that the arrival of particles in a direction
perpendicular to the nominal field occurred simultaneously or very shortly after the arrival of the first particles at L1 with $\mu$$\sim$1.  
Despite the inadequacy of the SOHO/ERNE field of view to apply the VDA method, we show its result in Figure 6d (dashed purple line).
It should be kept in mind that the inferred value of the parameter $A$=470$\pm$9 min can only be associated with the actual release time of the 
protons observed by SOHO/ERNE under the assumptions that all 
protons were released simultaneously and that their propagation up to the SOHO/ERNE telescope 
(including transport along field lines and acquisition of propagation direction perpendicular to the nominal field) depended uniquely on their speed and 
occurred on a common path length for all particles of different energies.
This common ``release" time $A$=470$\pm$9 min translates into 07:58 UT $\pm$9 min when we add  $\sim$8 minutes to compare with  remote-sensing observations
and should be considered as an upper limit of the release of the first particles directed toward L1.

We have added in Figure~6d (indicated by black dots) the onset times of near-relativistic electrons measured by WIND/3DP/SST in the pitch-angle bin
that first detected an intensity enhancement above the pre-event background intensity.
These onset times were obtained using near-relativistic electron measurements previously corrected 
for the contamination produced by $\sim$15$\%$ of incident electrons that scatter out of the silicon detector of the SST telescope
following the approach described by \citet{wang09}.
A least-squares fit to the WIND/3DP/SST electron onset times (dashed black line in Figure~6d) gives an estimated release time 
$A$=457$\pm$5 min (i.e. 07:45 UT $\pm$5 min again adding $\sim$8 min to compare with remote-sensing observations).
 We have also added  (indicated by a red dot) the onset time of the electron intensities measured in the sunward direction 
of the highest electron energy channel (175-315 keV) of ACE/EPAM/DE
(because of instrument response issues \citep[e.g.,][]{haggerty03} we consider only the use of this energy channel).
Note the similar onset times for electrons of comparable $1/\beta$ as observed by WIND/3DP and ACE/EPAM/DE.
We have also indicated (blue dot) the onset time detected in the  lowest electron energy channel of SOHO/EPHIN (0.25-0.70 MeV) that
occurred much earlier than the near-relativistic electrons.
By just time-shifting the onset time of the electron intensity enhancement along the nominal Parker spiral magnetic field line length $L_{Earth}$, 
we estimate a release time for the relativistic (0.25-0.70 MeV) electrons of  
07:35$\pm$2 min UT and for the near-relativistic (175-315 keV) electrons 
of 07:47$\pm$2 min UT (with the $\sim$8 minutes already added to allow comparison with remote-sensing observations).

Figure~6d also shows that there is a difference of $\sim$4 minutes between the onset times
of protons and electrons of similar speed (see onset times for the highest energy of SOHO/ERNE protons and 
the lowest energy of WIND/3DP electrons at $\beta$$\sim$1/2.7 in Figure~6d).
If both particle species were injected simultaneously and their propagation to L1 depended uniquely on their speeds, 
this discrepancy should be attributed to the different field of view of SOHO/ERNE with respect to WIND/3DP.
On the other hand, if proton and electrons were released at different times,  
as suggested in prior studies \citep[e.g.,][]{haggerty09}, a delay in the proton injection
with respect to electrons could contribute to the difference of $\sim$4 minutes observed between  the onset times of these electrons and protons.

\subsection{Solar observations}

The origin of the SEP event has been associated with   
a M6/3B flare from the NOAA Active Region 11719 at heliographic latitude +9$^{\circ}$ and longitude -12$^{\circ}$ as seen from Earth (i.e., N09E12)
with soft X-ray emission starting at 06:55 UT and peaking at 07:16 UT \citep[e.g.,][]{richardson14}.
Figure~7 shows from top to bottom, (a)  one-minute averages of  the soft X-ray intensities measured by GOES-15 and the intensities of the radio emissions 
as measured by (b) the S/WAVES detector \citep{bougeret08} on STEREO-B,
(c) the WAVES experiment \citep{bougeret95} on WIND, and
(d)  the S/WAVES detector \citep{bougeret08} on STEREO-A.
Both STEREO-B and WIND observed type III bursts starting from the highest frequencies that the instruments can detect ($\sim$16 MHz)
beginning at $\sim$06:58 UT, whereas for STEREO-A the Type III burst was observed
only at frequencies below 1 MHz starting around $\sim$07:00 UT.
A type II burst was observed by both STEREO-B and WIND starting
at 07:10 UT and ending at $\sim$15:00 UT in the range of
frequencies 10000-200 kHz (cf. lep694.gsfc.nasa.gov).
The low starting frequency of the Type III burst of STEREO-A indicates that the solar origin of this radio emission was obscured to this spacecraft,
and occurred on the occulted side of the Sun as seen from STEREO-A.
We note that from the onset time of the M6/3B flare up to the electron event
onset at STEREO-A, no significant
activity, that could be associated with the origin of the particle event at STEREO-A, 
was seen in the EUV and white-light images collected by both STEREO-A and STEREO-B.
Therefore, we associate the origin of the SEP event
at the three spacecraft with the activity related to the M6/3B flare and associated CME.

\subsubsection{EUV wave observations}

The location of the parent active region (cf. Figures~1~and~2) makes this event propitious to be observed by near-Earth spacecraft such as SDO.
We have used running difference images measured in the  193 {\AA}   and 211 {\AA} channels of 
SDO/AIA to characterize
the propagation of the associated EUV wave front.
 The left panel of Figure 8 shows a difference image computed from the 193 {\AA} intensity images
taken at 07:07 UT and 07:05 UT on 2013 April 11. The yellow contours visually identify the extent of the leading edge of the EUV disturbance
every two minutes (starting at 06:59 UT). 
The eastward section of the EUV wave was significantly brighter that the westward one. 
The eastward propagation was also faster based on the speed of the front projected on the plane of the sky. 
The wave front reached the footpoint of the field line connecting to STB estimated using the Parker spiral projection (cyan dot) at 
07:07 UT $\pm2$ min, 
whereas it reached the location of the STB footpoints using either the MAS (green diamonds) or the PFSS (red triangles) methods after 07:09 UT $\pm2$ min. 
The westward front  was faint and its propagation was not easy to track probably due to the presence of long-range loop systems connecting 
active regions. 
It is well-known that the presence of coronal structures, such as active regions and coronal holes, can complicate the tracking of EUV waves as they may partially reflect and refract the waves \citep[e.g.,][]{olmedo12}. 
Figure~2 shows the existence of large coronal loops (green field lines) a few degrees west of the central meridian line. 
The right panel of Figure~8 identifies the active regions on the solar disk just before the onset of the event. 
Two active regions (11718 and 11717) located near central meridian coincide with the series of loops and arcades shown in Figure~2 
and most likely were the structures preventing the wave to move westward. 
The wave front moving in that direction could not be seen after 7:15 UT, well before reaching the estimated footpoint of near-Earth observers, regardless of the model used to determine the location of this footpoint.

SDO/AIA 193 {\AA} and 211 {\AA} running-difference images (not shown here) clearly show that the  wave front
propagating eastward extended in the off-limb coronal portions of the images
(indicated by the most external yellow contours in the left panel of Figure~8).
This suggests that the wave front identified in Figure~8 may not result from a structure propagating
on the solar disk, but rather result from the projection of the expanding CME-driven shock
propagating at increasing altitudes in the solar corona.
Therefore, it is possible  that the observed  EUV wave resulted from projection effects of the flanks of the CME that propagated mostly eastward (as seen from Earth).
The portion of the CME-driven shock moving westward, i.e. toward the regions where L1 observers were magnetically connected,
propagated at high altitudes but 
did not leave any EUV trace on the solar disk (see Section 2.3.2).

Apart from visual identification of the EUV wave, we have also computed the EUV intensity increase along two arc sectors covering the region
from the source of the wave to the estimated location of the magnetic footpoints
of STEREO-B and L1. 
The technique used is similar to that described by
\citet{long11a, long11b}, \citet{muhr11} and \citet{prise14}. 
We calculate the intensity of percentage running-difference of 193 {\AA} images from SDO/AIA at different times over 
two 22.5$^{\circ}$-wide
arc sectors centered at the active region source of the wave and
including the footpoints of STB  and L1 identified using either the MAS or the PFSS model, respectively.
Figure~9 shows the intensity increase in the direction of the STEREO-B footpoint (top panel) and in the direction
of L1-observer's footpoint (lower panel) at different times along these two arc sectors.
The curves represent the average
intensity per pixel assuming a 1 degree latitude resolution from the location of the flare site.
The units in the vertical axis of the two panels indicate the factor by which the intensity increased in the selected wedge
(units used are consistent between the top and bottom panels, indicating that the EUV wave was more intense in its way to the STEREO-B footpoint).
We see that the EUV front in the wedge containing the footpoint connecting to near-Earth observers 
weakened considerably at around $\sim$07:13 UT, only able to reach a longitude $\sim$W25$^{\circ}$
(as seen from Earth) but never the magnetic connecting footpoint of the near-Earth observers.
By contrast, the EUV wave in the sector containing the footpoint of STEREO-B propagated without any hinderance
to extend even above the east limb of the Sun (as seen from Earth).
The estimated arrival time of the leading edge of the EUV wave front at the STEREO-B footpoint is 07:07$\pm$2 min for the case estimated using the Parker Spiral 
method, and 
07:11$\pm$2 min for the case estimated using either the MAS or the PFSS methods.
Note that the location of the STEREO-B footpoints close to the east limb (as seen from SDO) introduces
a certain ambiguity in the exact identification of the EUV wave in terms of the distance traveled above the solar surface.

\subsubsection{White-light coronagraph observations}
 
The CME associated with the origin of the SEP event was first observed by the LASCO C2 coronagraph  at 07:24 UT at 2.79 R$_{\odot}$ propagating on the eastern limb
as seen from an Earth's centered observer (Position Angle = PA = 85$^{\circ}$) with a plane-of-sky speed of  861 km s$^{-1}$ (cdaw.gsfc.nasa.gov/CME$\_$list/).
The CME was also seen erupting off the western limb of STEREO-B at 07:10 UT above 1.8 R$_{\odot}$ (with a sky-plane speed of 1388 km s$^{-1}$ 
in its leading edge according to secchi.nrl.navy.mil/cactus/).
From the STEREO-A perspective it was a backside event that became a halo CME at around $\sim$07:40 UT.

Recently, \citet{kwon14} developed a compound geometrical model to determine the 3D structure
associated with CMEs using data from the two STEREOs, SDO and SOHO.
An ellipsoid shape centered at a certain altitude $h_{E}$ is used to describe the outermost front driven by the CME.
The CME ejecta (i.e., the bright frontal loop or three part CME structure) is described by 
 the Graduated Cylindrical Shell (GCS) model developed by 
\citet{thernisien09} and \citet{thernisien11}. 
 The
 fitting is done via a forward modeling approach. Namely,  a set of images of
the event obtained from the three viewpoints combining data from STEREO/SECCHI/EUVI \citep{wuelser04}, STEREO/SECCHI/COR \citep{howard08}, SDO/AIA \citep{lemen12} and  SOHO/LASCO \citep{brueckner95} is selected. 
The images should be as co-temporal as possible to minimize evolutionary effects on the
event morphology. Then, we assume the geometric shape of the structure to be fitted; GCS
for the magnetic flux rope like structure, and the ellipsoid for the shock front. The structure is then
projected onto the three images, taking into account the viewing geometry and is manipulated
until a visually satisfying fit is obtained in all viewpoints simultaneously. The
GCS model has 7 free parameters, such as height, longitude, and latitude of the center of the flux rope,  as well as the half angle, aspect ratio, height and rotation of the flux rope, as described in detail by \citet{thernisien11}.
The ellipsoid model uses also 7 free parameters, such as height, longitude and latitude of the center of the ellipsoid, the length of the three semi-principal axes,
and rotation angle of the ellipsoid 
\citep[see details in][]{kwon14}.

Figure~10 shows, for different times, representations of the outermost front (using the ellipsoid model) 
 overplotted in the observed images
(for clarity purposes the modeled GCS flux rope is not represented).
The outermost front is 
shown by red, orange, blue and cyan colors (each color for each quadrant).
The white circle identifies the maximum extension of the outermost shock front. 
Dashed lines are used when the reconstructed structure is on the other side of the plane of the image.
Through iterative fittings, we find that the outermost front is reproduced well with a sphere.
At the beginning of the eruption (until time 07:10$\pm$5 min UT) the intersection of the ellipsoid
with the solar surface spatially coincided with the EUV wave front,
for later times (after 07:25$\pm$5 min UT) the associated shock started to expand above western
longitudes (as seen from Earth) without any signature left close to the solar surface.
Eventually, at time  07:40$\pm$5 min UT the ellipsoid became large enough, that it did not intercept with the solar surface,
but enclosed the whole solar corona.

The right panels of Figure~11 summarize the determined 3D morphology of the outermost front (black lines) 
at different times in a 
selected plane, y'-z' plane, as shown in the left panels of Figure~11. 
The plane is defined from three points on the solar surface: the center of the ellipsoid projected radially on the solar surface and the magnetic footpoints of STEREO-B and L1 
estimated using the Parker Spiral method (top) and the MAS method (bottom).
In Figures 11b and 11d, the black dashed lines in the outermost front at distances below 1.5 R$_{\odot}$ (i.e., the inner boundary of the STEREO/SECCHI/COR-1 coronagraph)
are used to indicate the lower portion of the shock front for those times when the EUV wave front cannot be tracked.   
It is found that the lower portion of the shock front moving eastward as seen from Earth  (i.e. the left flank of the shock in Figure~11) 
sweeps through the magnetic footpoints of STEREO-B estimated using the 
Parker Spiral method and the MAS method at 07:15$\pm$3 min UT and 07:21$\pm$4 min UT,  respectively.
The errors associated with these times are estimated considering a 10$\%$ uncertainty in the geometry of the modeled shock front.
The discrepancy between these times and those obtained in Section 2.3.1 based on the tracking of the EUV wave front as seen from SDO/AIA
may result from projection effects. Figure~12 shows the estimated shape of the shock front at 07:07 UT as estimated from the ellipsoid-GCS models
(the two curved gray lines around the front shock indicate the 10$\%$ uncertainty in the geometry of the modeled shock front).
The solid straight line indicates the line of sight from a near-Earth observer (such as SDO).
The projection of the shock front over the solar surface
may suggest that the EUV wave is already passing over the STEREO-B foopoint at this time.
However, the actual footprint or ``ground track" of the 3D wave on the solar surface may be away from the projected EUV wave front.

The prolongation of the west flank of the 
shock front above the solar surface, allows us to determine the passage of its footprint above the magnetic footpoints of L1 which is found  
to be 07:23$\pm$4 min UT (using both MAS and Parker spiral projections). However, the non-observation of the EUV front on this point
suggests that the shock wave driven by the CME was located at higher altitudes not being capable of  leaving an observable EUV trace on the solar surface.
Therefore, this time only refers to the passage of the prolongation of the right flank of the shock in Figure~11
but not the actual passage of the lower portion  of the shock over the solar surface. 
  
 Tracking the shock front at different times allows us to estimate its location at the times we inferred for the release of particles observed at STEREO-B and L1.
 Table~2 lists the estimated SEP release times using the different methods described in Section 2.2 together with the height of the shock front 
 above the magnetic footpoint of each spacecraft at the SEP release times.
The errors associated with the shock heights  are estimated
considering a 10$\%$ uncertainty in the geometry of the modeled shock front.
At the estimated release time of the protons observed by STEREO-B (07:10$\pm$4 min UT using the VDA method),
the shock
front only reached the STEREO-B magnetic footpoint computed by the Parker spiral method within the error bars
(07:15$\pm$3 min UT), hence the height is given only as an upper limit.
For the rest of heights above STEREO-B footpoint we use the location based on the Parker spiral method. 
We see that the release of particles observed at L1 occurred when the portion of the shock above the footpoint was at higher altitudes than the portion of the shock above the STEREO-B footpoint when particles observed by this spacecraft were released.

An important aspect in terms of particle acceleration at shocks is the angle between the normal to the shock and the local magnetic field
\citep[e.g.,][]{giacalone06, battarbee13}.
The reconstruction of the shock front allows us to determine the normal to the shock and the angle formed between this normal and the radial direction ($\theta_{nr}$).
Column 4 of Table~2 lists the values of $\theta_{nr}$ at the time indicated in Column 2, at the distance indicated in Column 3 above the footpoint of the observer identified in Column 1 of Table~2.
The low altitude of the portion of the shock (i.e. its east flank)
moving over the STEREO-B footpoint at the time of the particle release yields values of $\theta_{nr}$
close to $\sim$90$^{\circ}$. By contrast, the portion of the shock moving over the L1 footpoint was already located at high altitudes
at the time estimated for particle release, yielding smaller  values of $\theta_{nr}$. Note that these altitudes are above the PFSS source region (i.e.
above the heliocentric distance 2.5 R$_{\odot}$) where open field lines provided by PFSS are already radial.
The fact that both MAS and PFSS models give the same location for the footpoint of near-Earth observers (Table~1)
indicates that the approximation of radial open magnetic fields above R=2.5 R$_{\odot}$ is also validated in the MAS model.
Therefore, based on the results of the fitted shock, the flank of the shock moving over the STEREO-B footpoint at the time of the SEP release can be considered,
in a good approximation, quasi-perpendicular; whereas the portion of the shock above the L1 footpoint
at the time of the SEP release in that direction can be considered quasi-parallel.

 \section{Discussion}
 
 The multi-spacecraft SEP event on 2013 April 11 was generated by a single eruption originated from the NOAA active region 11719 at N09E12.
 The site of this eruption was well separated from the estimated footpoint locations of STEREO-A, STEREO-B and L1 observers.
 However, an intense SEP event was observed by both STEREO-B and L1 spacecraft 
 (whose footpoints were separated by $\sim$60$^{\circ}$ and $\sim$70$^{\circ}$ in longitude from the 
 site of the parent active region, respectively), whereas STEREO-A 
 (whose footpoint was around $\sim$170$^{\circ}$ from the eruptive source region) detected only a significant electron event, but no increases in proton or heavy ion intensities. 
 
As suggested by previous works (see Section 1),
we have examined the possibility that the extent of the EUV disturbance in the solar corona was associated 
with the extent of the expanding coronal
 shock able to inject particles into the interplanetary medium and hence with 
 the longitudinal extent of the SEP event in the heliosphere.  
 The tracking of the EUV wave front using SDO observations  
 shows that it arrived at the STEREO-B footpoint between 07:07 UT and 07:11 UT (depending on the method used to estimate the footpoint location),
 but it never reached the region where magnetic field lines connect near-Earth observers to the Sun.
 The observation of an intense SEP event  near-Earth   implies that  
the extent of the EUV wave cannot be used reliably as a proxy for the longitudinal extension of the SEP events in the heliosphere.

 Since EUV waves are 3D phenomena and not just surface disturbances
 \citep[e.g.,][]{patsourakos09, veronig10, kwon13},
we studied the full 3D structure associated with the CME as seen in white-light and EUV images.
 Projection effects of this 3D structure on the solar disk may lead to errors in the estimated arrival time  of the EUV wave front 
 at the  location of the magnetic footpoints connecting to each spacecraft.
Our technique, based on the model by \citet{kwon14}, allows us to improve the reliability of the timing between the shock arrival at the
magnetic footpoints and the particle release times.
 Whereas
\citet{rouillard12} considered simpler representations of the expanding CME shock and EUV wave
(such as symmetric wave propagation) and
relied on projection on synoptic maps to extract positional information for the shock,
our approach fits simultaneously EUV and coronagraph observations, and allows us to obtain the 3D time history 
and location of the shock driven by the CME, in particular at the times estimated for the release of particles.
The shock propagated faster toward the STEREO-B footpoint  than toward the L1 footpoint site.
Consequently, longitudinal distributions of peak intensities and fluences observed for this event \citep[e.g.,][]{richardson14, cohen14}
show an offset toward the east (i.e. toward STEREO-B) with respect to those averaged distributions
obtained over a large number of events \citep[e.g.,][]{lario13}.

It is noteworthy that the methods used to determine the SEP release times
(either velocity dispersion or time-shifted analyses) are not free of uncertainties.
\citet{vainio13} reviewed the success and reliability of these methods.
Uncertainties about (i) the actual path length and shape of the interplanetary magnetic field lines
along which SEPs propagate, (ii) the conditions
of SEP interplanetary transport, (iii) the actual injection profile of particles of different energies, and (iv) the identification of the exact
event onset times pose serious difficulties in the use of these methods widely employed.
Additionally, the viewing direction of the particle instruments plays also a role in the detection of the first injected particles.
For these reasons, we have considered several delimiting values for the release time of the particles observed by STEREO-B and L1 observers
as listed in Table~2.
Within the  uncertainties, the arrival of the EUV wave front and of the structure associated with the CME-driven shock
at the footpoint of STEREO-B are consistent, within the uncertainties,  with the release
of particles toward this spacecraft. We cannot distinguish, however, whether there was an actual delay in the injection of electrons and protons observed by STEREO-B.
Similarly, the orientation of the particle instruments on SOHO
does not allow us to determine whether the release of protons was delayed with respect to the electrons.

Table~2 also lists the height of the shocks above the footpoints at the estimated particle release times. 
Several studies have indicated that SEP injection starts when the associated CME is already high in
the corona. For example, estimates of the release time of both near-relativistic ($>$30 keV) electrons and
high-energy (470 MeV-4 GeV) protons observed in Ground-Level Events (GLEs) showed that,
when the particle injection starts, the leading edge of the CMEs can already be at heliocentric distances $>$2 R$_{\odot}$ for electrons \citep{simnett02}
and between 2 and 7 R$_{\odot}$ for GLEs \citep{reames09}. Similarly, at the time of the
observed peak intensity in GLEs, CMEs can already be at 5-15 R$_{\odot}$ or even greater heliocentric
distances \citep{kahler94}.

At the estimated release time of the particles observed by STEREO-B, the leading-edge of the CME was starting to show up in the images 
taken by the STEREO-B/SECCHI/COR-1 coronagraph  (Figure~10a).
However, the modeling of the propagation of the shock wave shows that 
the portion of the CME-driven shock above the STEREO-B magnetic footpoint (i.e. its left flank as shown in Figure~11) was moving very close to
the solar surface. Assuming that open magnetic field lines tend to be radial at these low altitudes, we estimate that the flank of the shock
moving over STEREO-B footpoint was quasi-perpendicular and therefore capable of accelerating ions and electrons at high energies provided that 
an energetic seed population exists \citep{giacalone06}.

At the estimated release time of the particles observed at L1, the leading-edge  of the CME was already at $\sim$2-4 R$_{\odot}$ above the solar surface 
as seen in SOHO/LASCO/C2 and STEREO-B/SECCHI/COR-2 coronagraph images. 
Naturally, the portion of the shock moving above the footpoint of near-Earth observers (the right flank of the shock as shown in Figure~11)
was located at lower altitudes than its leading edge but
it did not leave any EUV signature on the solar disk.
In agreement  
with \citet{prise14}, we suggest  that the particles observed 
by SOHO were accelerated by the CME-driven shock  at high altitudes but this portion of the shock
was not observed in the  EUV images. 
Assuming that open magnetic field lines were radial at these altitudes, we estimate that the portion of the shock
moving over L1 footpoint was more quasi-parallel than over STEREO-B. 
However, both locations registered  Fe-rich SEP events with high-energy $\sim$100 MeV protons and relativistic $\sim$1 MeV electrons.

Finally, the observation of electrons by the poorly connected STEREO-A spacecraft suggests
that these fast particles (compared to energetic protons) were able to spread in longitude more efficiently
than protons and heavy ions. 
The long delay observed at the onset of the event at STEREO-A and the small anisotropy detected by this spacecraft indicate that these particles 
were not directly injected onto the field lines connecting STEREO-A with the Sun, but rather  some transport process in the interplanetary medium
allowed them to reach STEREO-A.
Note that although the modeled CME-driven shock encircled the Sun at about $\sim$07:40 UT,
the near-relativistic electron enhancement at this spacecraft was not observed until
$\sim$11:50 UT.
The lack of type II emission observed by STEREO-A (Fig.~7) suggests that the portion of the shock
covering the opposite side of the Sun (as seen from Earth)
was too weak to accelerate electrons.
The fact that electrons can be observed from poorly connected longitudes suggests that  electron events should be more commonly
observed regardless of the longitudinal separation between parent active regions and the spacecraft.
In fact, the longitudinal distributions of SEP peak intensities in events simultaneously observed by several spacecraft
tend to be wider for near-relativistic electron intensities than for energetic proton intensities
\citep[e.g.,][]{lario13}.

\section{Summary}

The detailed analysis of the SEP event on 2013 April 11 allows us to conclude that 
(i) SEP events can be observed by spacecraft even when the associated EUV wave front does not reach the magnetic connection footpoint of such spacecraft;
(ii) near-relativistic electrons spread over wide range of heliolongitudes allowing the observation of gradual electron intensity increases even at poorly connected spacecraft;
(iii)  release of energetic particles for poorly connected spacecraft may occur when the associated CME-driven shock is already high in the corona without leaving any signature in the EUV observations low in the corona; and (iv) projection effects of the 3D structure associated with the parent CME may lead to
erroneous identification of the arrival time of EUV wave fronts
at the footpoints of the field lines
connecting to well-separated spacecraft.
For these reasons, we conclude that EUV waves cannot be used reliably as proxy for the broad longitudinal extent of SEP events;
this does not exclude the possibility of this assumption to work in some other cases.

\acknowledgments
We acknowledge the STEREO PLASTIC, IMPACT, SECCHI, SWAVES; SOHO LASCO,  EIT, ERNE, EPHIN; ACE EPAM, MAG;  WIND 
3DP, MFI, WAVES; and SDO AIA, HMI  teams for providing the data used in this paper. 
The SDO/AIA and SDO/HMI data are provided by the Joint Science Operations Center (JSOC) Science Data
Processing (SDP). SDO is the first mission to be launched for NASAÕs Living With a Star (LWS) Program.
The STEREO/SECCHI data are produced by an international consortium of the NRL (USA), LMSAL (USA), NASA-GSFC (USA), RAL (UK), University of Birmingham (UK), MPS (Germany), CSL (Belgium), IOTA (France), and IAS ( France).
SOHO is a mission of international cooperation between ESA and NASA.
The ACE data are provided by the  ACE Science Center and the  WIND data by the Coordinated Data Analysis Web.
The present work benefited from discussions held at the International Space Science Institute (ISSI, Bern, Switzerland) within the frame 
of the international team ``Exploration of the
inner Heliosphere what we have learned from Helios and what we want to study with
Solar Orbiter", led by Dr. W. Droege.
DL acknowledges the support from NASA under grant NNX11A083G and the ACE grant NNX10AT75G.
JZ and R.-Y. K. acknowledge the support from NSF grants ATM-0748003, AGS-1156120 and AGS-1249270.
RGH acknowledges the financial support by the Spanish MINECO under project AYA2012-39810-C02-01.

\clearpage



\begin{figure}
\epsscale{.40}
\plotone{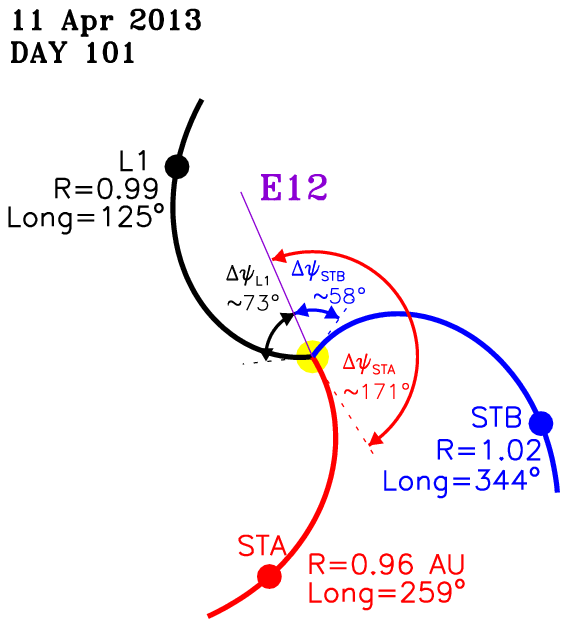}
\caption{View from the north ecliptic pole showing the location of STEREO-A (STA; red symbol), near-Earth observers (L1; black symbol) and STEREO-B (STB; blue symbol) 
on day 101 of 2013 (2013 April 11). $R$ and $Long$ indicate  the heliocentric distance and heliographic inertial longitude of each observer, respectively.
Also shown are nominal interplanetary magnetic field lines connecting each spacecraft with the Sun (yellow circle at the center) considering the 
solar wind measured at the onset of the SEP event.
The purple line indicates the longitude of the parent active region (E12 as seen from Earth, E149 as seen from STEREO-A, and W129 as seen from STEREO-B) and
the angles $\Delta{\psi}$ indicate the longitudinal distance between the active region and the footpoints of the nominal Parker spiral field lines.
(A color version of this figure is available in the online journal). \label{fig1}
}
\end{figure}

\clearpage

\begin{figure}
\epsscale{.80}
\plotone{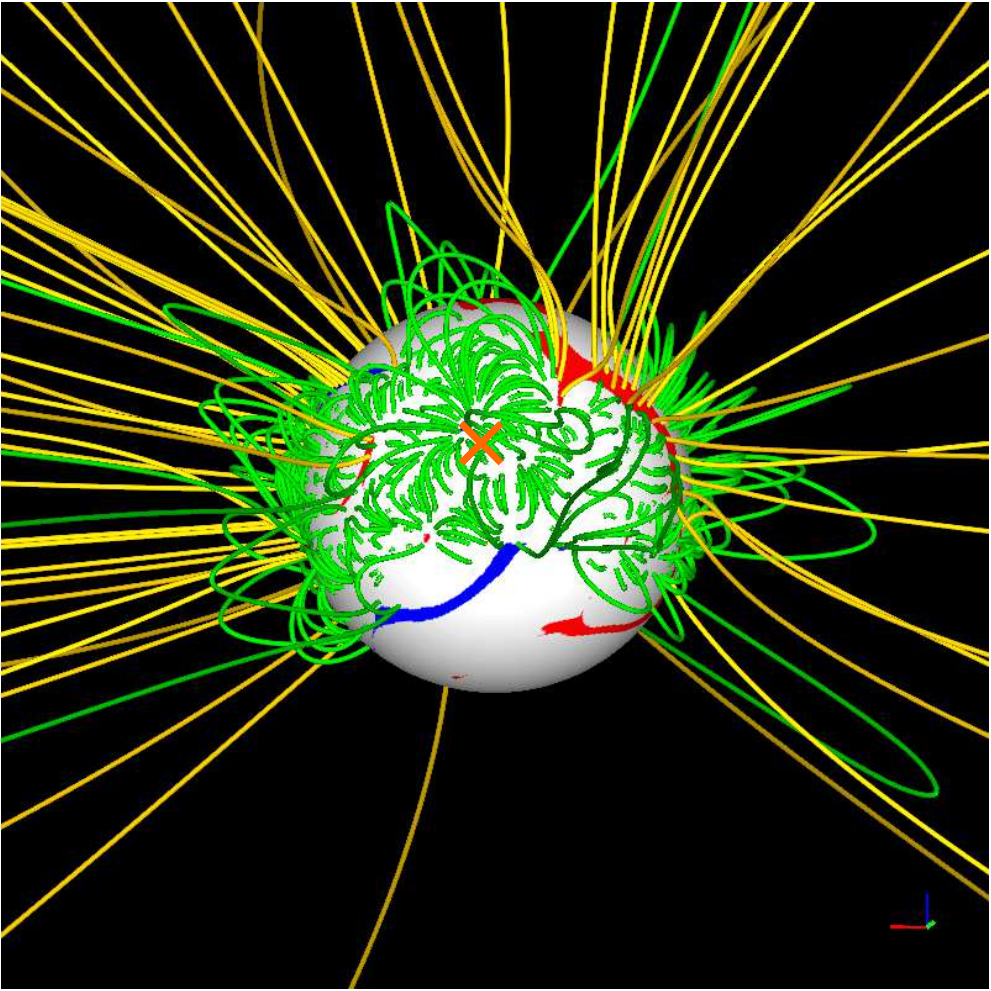}
\caption{Model view of the Sun from Earth on 2013 April 11 at 06:04 UT.
The red (positive) and blue (negative) patches on the solar surface (at 1 R$_{\odot}$) show the location and polarity
of the coronal holes. 
A selection of closed (green) and open (gold) field lines are shown.
The orange cross indicates the location of the active region where the M6/3B flare occurred.
(A color version of this figure is available in the online journal).
\label{fig2}}
\end{figure}

\begin{figure}
\epsscale{.80}
\plotone{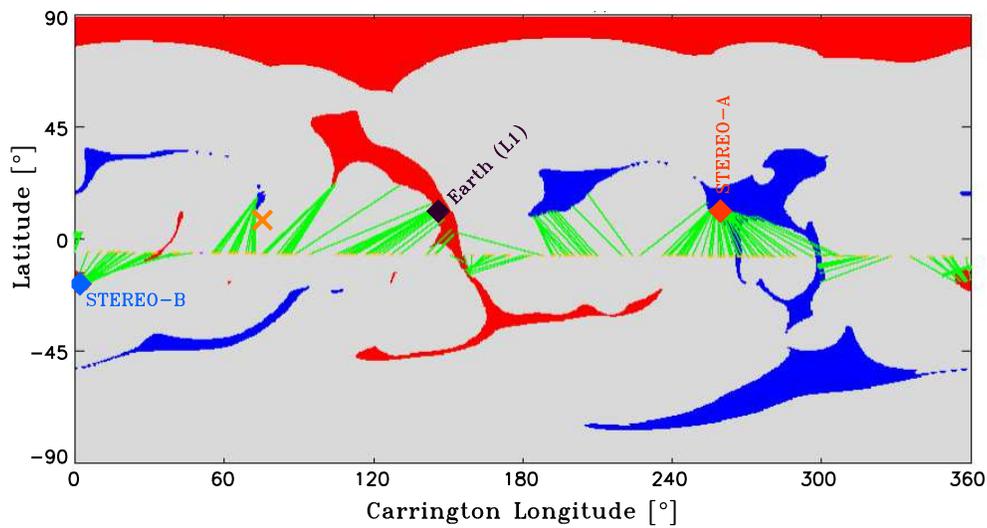}
\caption{Coronal holes computed from the MAS simulation color-coded according to the observed underlying
photospheric magnetic field (red for outward and blue for inward magnetic field 
polarity).
Earth's trajectory is superimposed (thin orange dots), together with
the mapped source regions of the plasma measured at ACE (indicated by the green lines).
Earth was at Carrington Longitude 85$^{\circ}$ at 07:00 UT
on 2103 April 11.
The black diamond identifies the footpoint of the magnetic field line connecting to ACE
at that time. We have also indicated with blue and red diamonds the location of the footpoints of the field lines 
connecting to STEREO-B and STEREO-A, respectively. The orange cross identifies the site of the parent flare. 
(A color version of this figure is available in the online journal).
\label{fig3}}
\end{figure}

\begin{figure}
\epsscale{.80}
\plotone{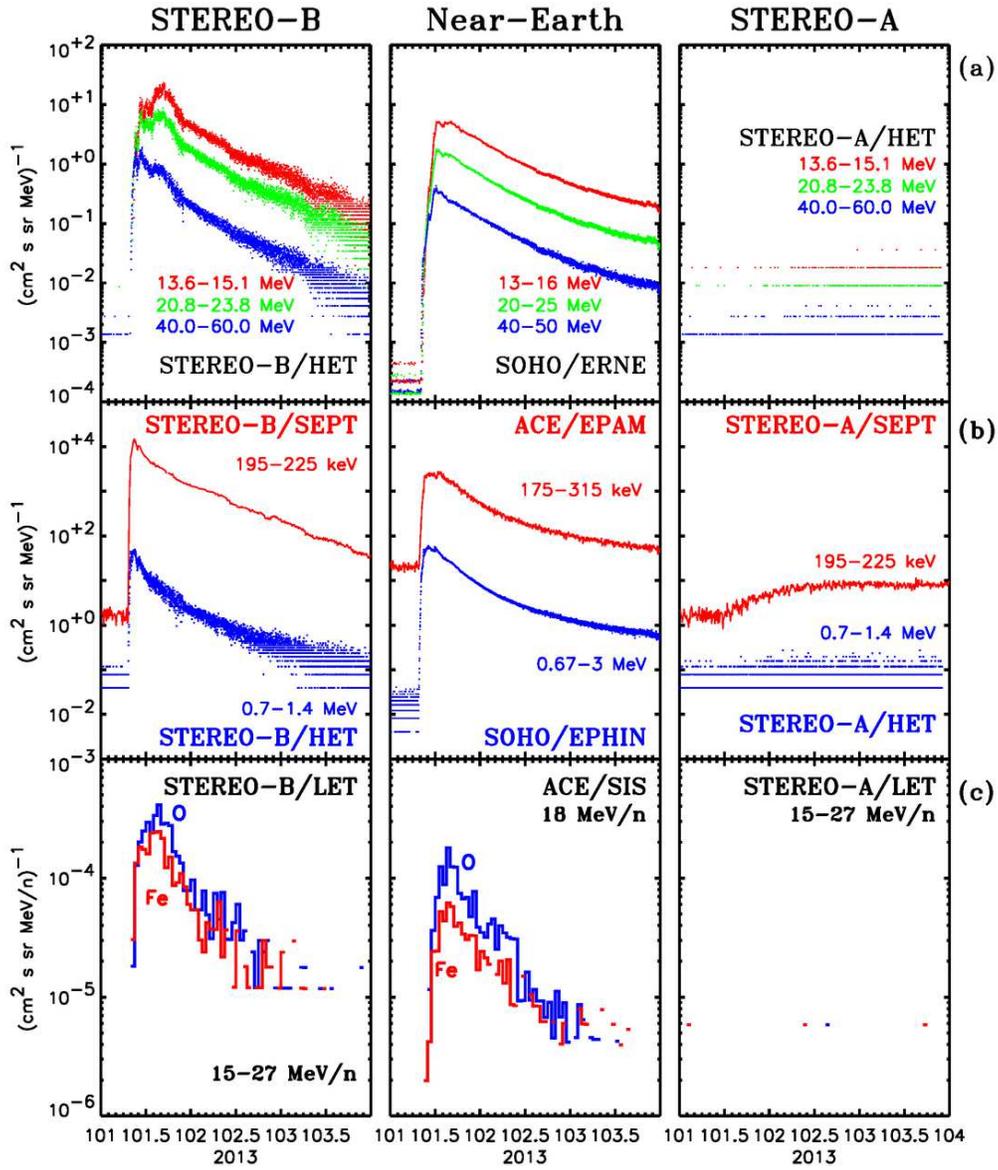}
\caption{Intensity-time profiles of, from top to bottom, protons, electrons and heavy ions,
as measured by, from left to right, STEREO-B, L1 observers and STEREO-A. 
(A color version of this figure is available in the online journal).
\label{fig4}}
\end{figure}

\begin{figure}
\epsscale{1.1}
\plotone{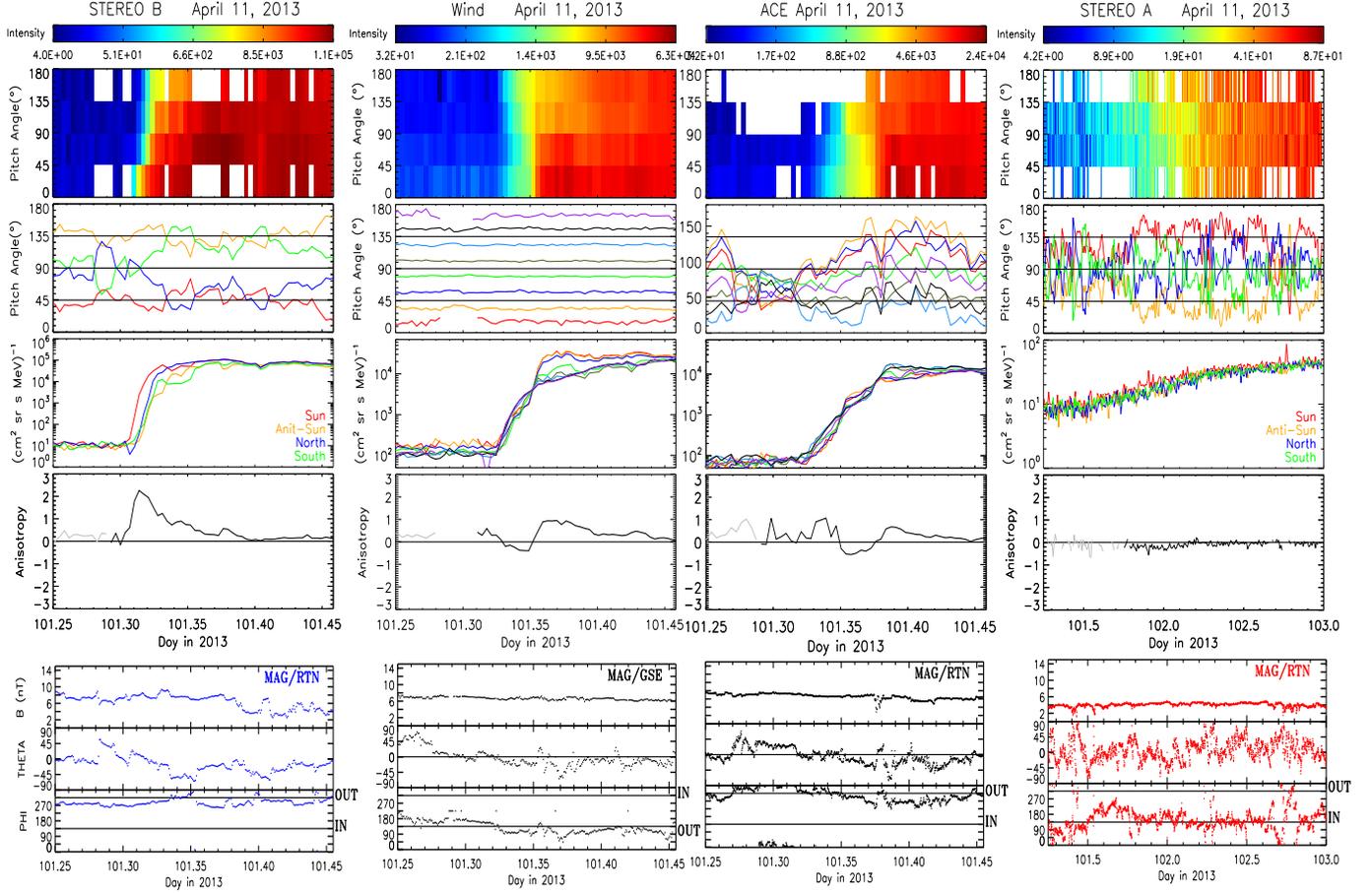}
\caption{Anisotropy and intensity time profiles of near-relativistic electrons as observed by, from left to right, STEREO-B,
WIND, ACE, and STEREO-A. From top to bottom: Pitch-angle dependent intensity distribution
color coded according to the intensity indicated in the top horizontal color bar
(units are electrons cm$^{-2}$ sr$^{-1}$ s$^{-1}$ MeV$^{-1}$).
Pitch-angle direction measured at the center of the four telescopes of STEREO/SEPT
(Sun(red), Anti-Sun (orange), North (blue), and South (green)), of the 8 pitch-angle bins of WIND/3DP/SST, and of the 8 sectors
of ACE/EPAM/LEFS60.
Intensity measured in each one of the 4 telescopes of STEREO/SEPT, each one of the 8 pitch-angle bins of WIND/3DP/SST, and each one of the 8 sectors of 
ACE/EPAM/LEFS60.
First-order anisotropy $A$ (the periods indicated with dotted gray traces correspond to periods with intensities
close to background intensities).
Magnitude and angular coordinates of the interplanetary magnetic field (RTN coordinate system is used for STEREO-A, STEREO-B and ACE, and GSE for WIND). 
The energies covered are 55-105 keV for the two STEREOs, 58-82 keV for WIND, and 62-102 keV for ACE. 
(A color version of this figure is available in the online journal).
\label{fig5}}
\end{figure}

\begin{figure}
\epsscale{.62}
\vspace{-0.7cm}
\plotone{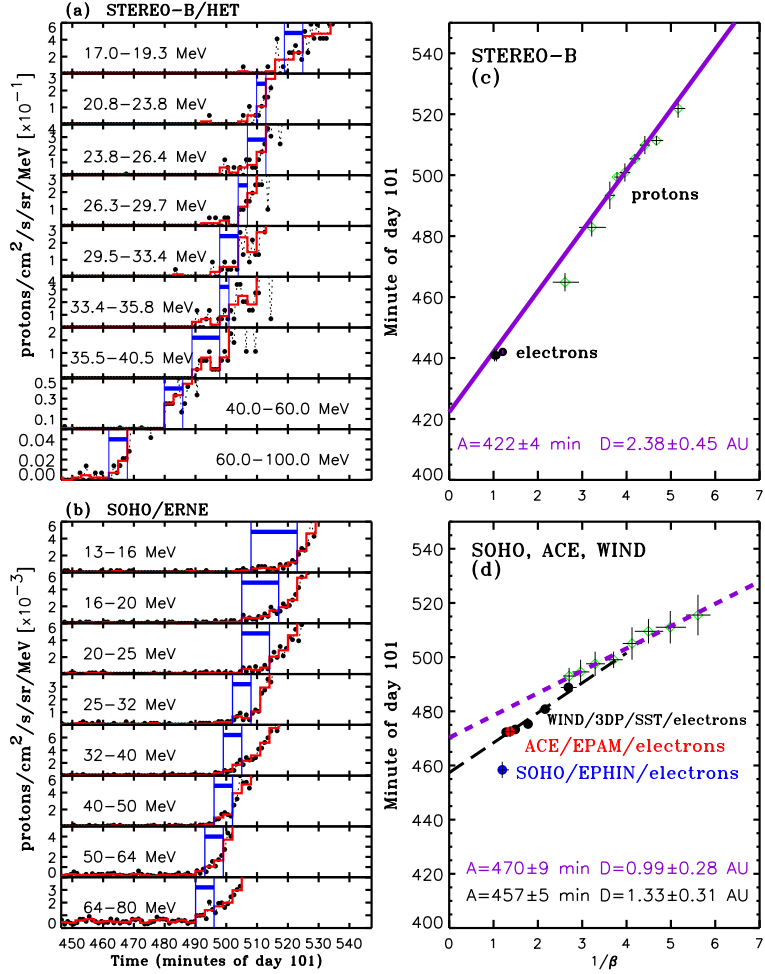}
\caption{(Left) Onset of the SEP event as observed by (a) the proton energy channels of STEREO-B/HET
and (b) the proton channels of SOHO/ERNE/HED.
The black dots are one-minute averages, whereas the red line are three-minute averages.
The blue  vertical lines indicate the time interval when the onset of the event is identified.
(Right) Velocity dispersion analysis
of the onset of the event
at (c) STEREO-B and (d) L1.
The green symbols identify the proton onset event identified in the left panels. 
The black symbols in panel (c) identify the onset of the event as observed by the 375-425 keV electron channel of STEREO-B/SEPT and the 0.7-1.4 MeV electron channel of STEREO-B/HET. The black symbols in panel (d) identify the onset of the event as observed in different energy channels of WIND/3DP/SST,
the red symbols the onset in the 175-315 keV electron channel of ACE/EPAM/DE and the blue symbol in the 0.25-0.70 MeV electron channel of SOHO/EPHIN. 
The purple straight lines are linear regression fits to all proton data points and the dashed black line to WIND/3DP electron data points.  
The legend gives the estimated release time ($A$) and the path length ($D$) discussed in the text. 
(A color version of this figure is available in the online journal).
\label{fig6}}
\end{figure}

\begin{figure}
\epsscale{.70}
\plotone{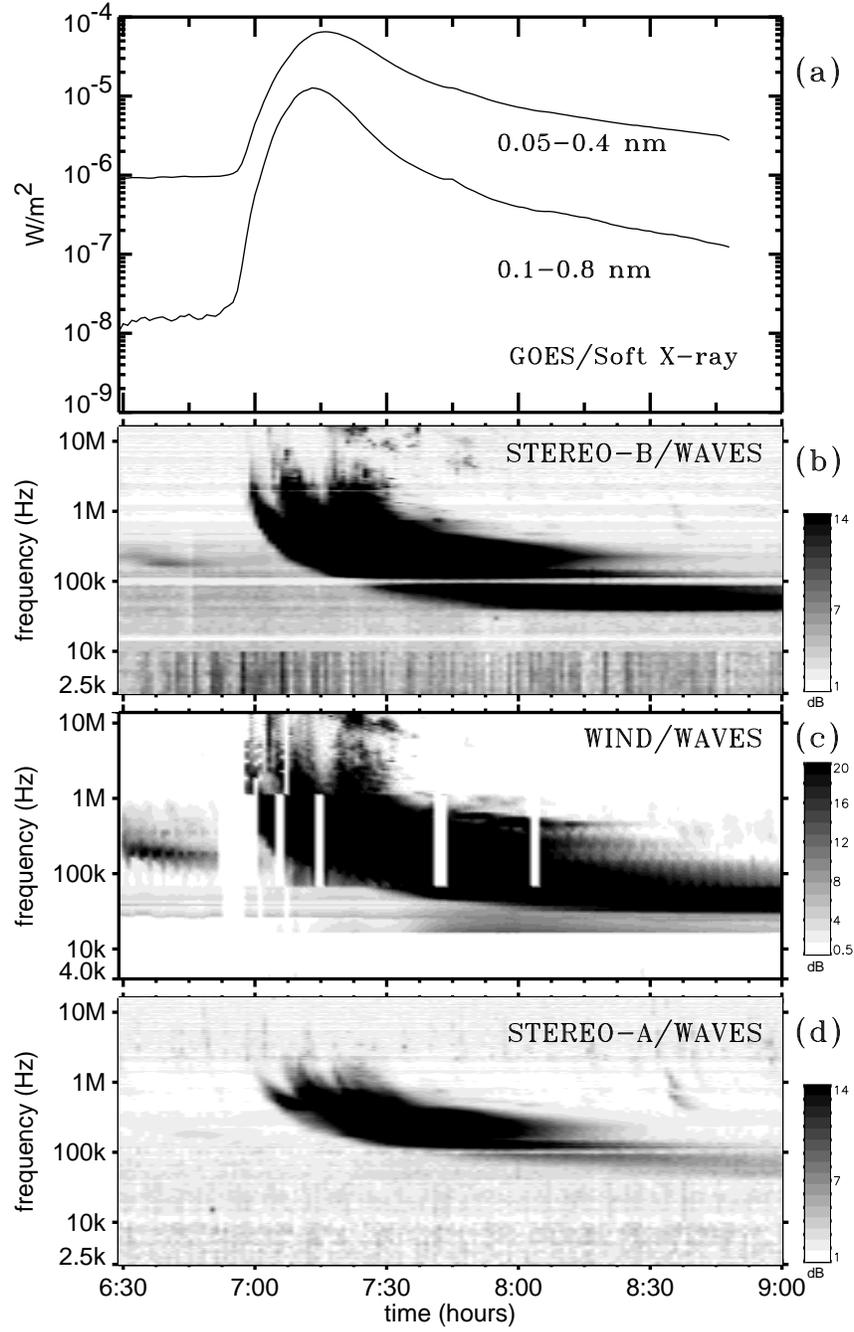}
\caption{(a) Soft X-ray GOES intensities. Radio measurements from (b) SWAVES on STEREO-B,
(c) WAVES on WIND, and (d) SWAVES on STEREO-A. \label{fig7}}
\end{figure}

\begin{figure}
\epsscale{1.0}
\plotone{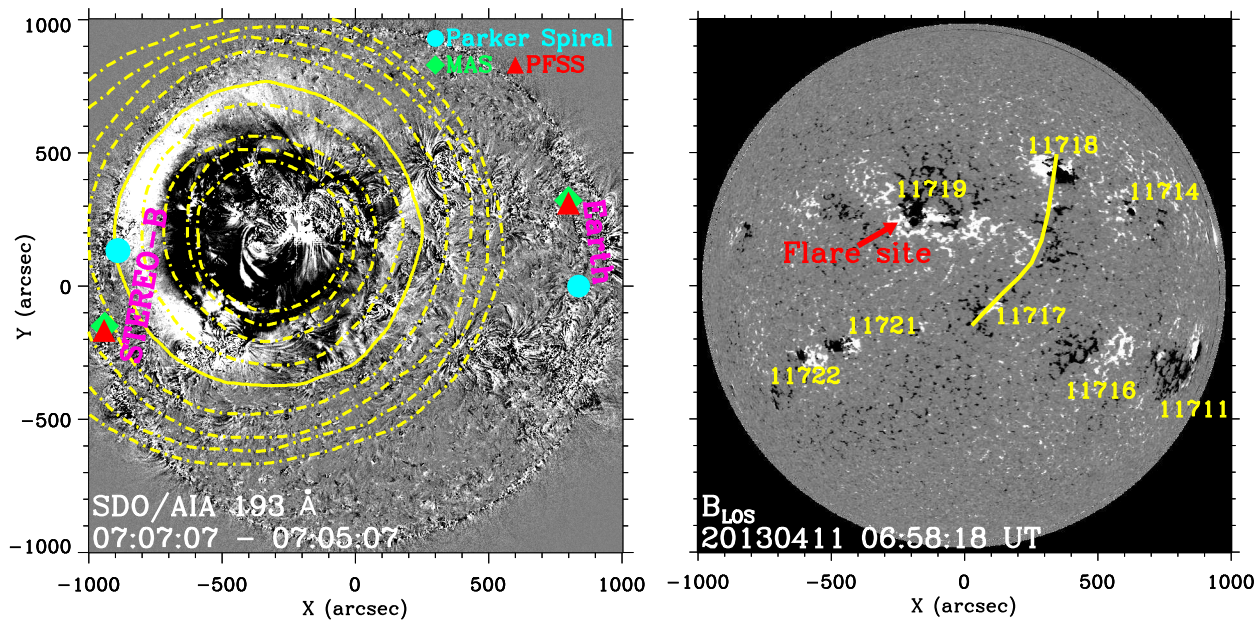}
\caption{(Left) Difference image of the SDO/AIA 193 {\AA} images obtained at 07:07 UT and 07:05 UT on 2013 April 11.
The yellow contours identify locations at 2-min cadence of the EUV front starting at 06:59 UT.
The cyan, green and red symbols indicate the locations of the footpoints of the field lines connecting to STEREO-B (east limb) and Earth (west limb)
using the Parker Spiral, MAS and PFSS methods, respectively.
(Right) One-hour average HMI LOS magnetogram just before the occurrence of the parent solar flare.
The numbers indicate the different NOAA active regions.
The yellow curve indicates the neutral line marking the magnetic structure most likely responsible for preventing the westward propagation of the wave.
This structure coincides with the large-scale field lines shown in Figure~2.
The path toward the STEREO-B footpoint was relatively clean.
(A color version of this figure is available in the online journal).
\label{fig8}}
\end{figure}

\begin{figure}
\epsscale{.80}
\plotone{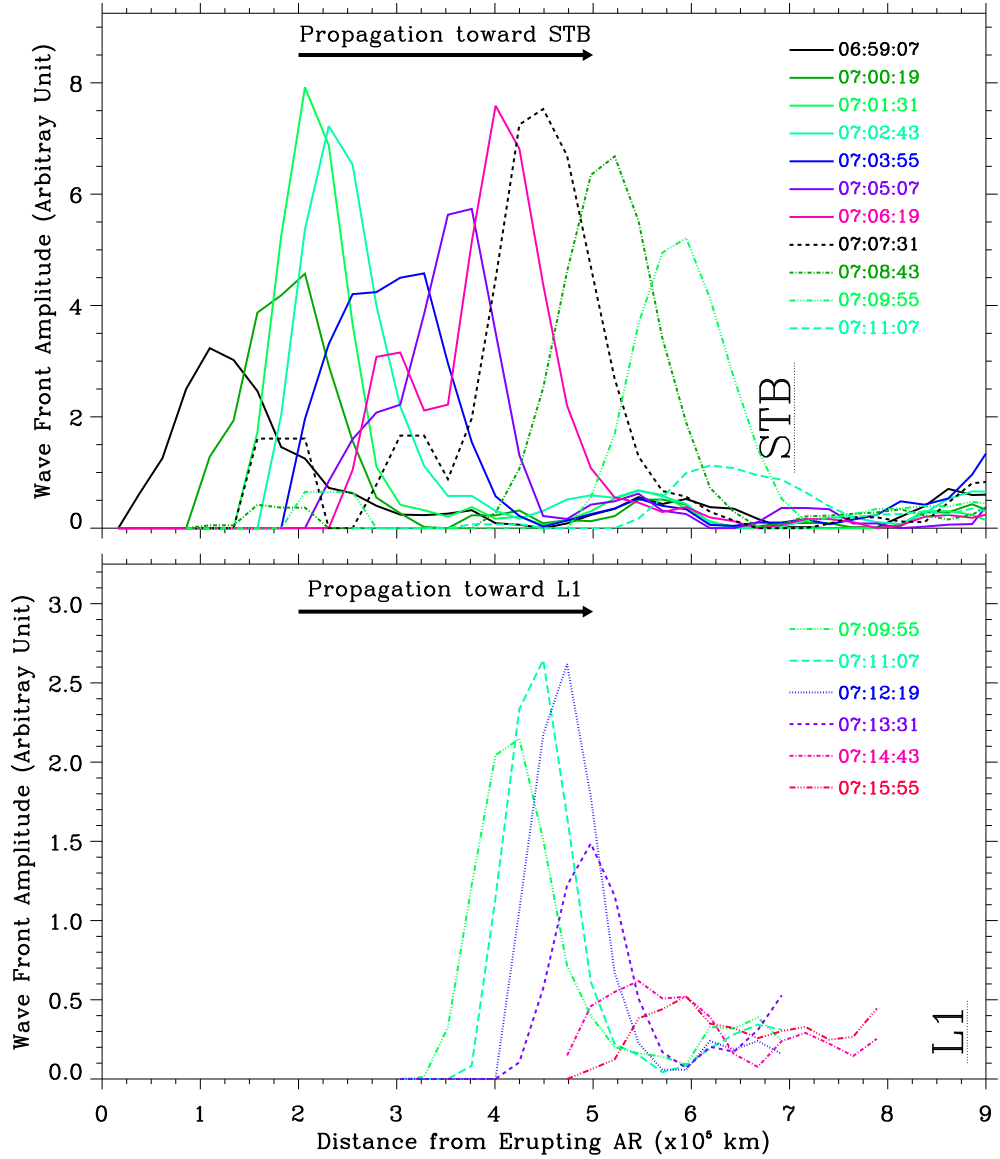}
\caption{EUVI intensity increase at different times over two sector wedges covering the distance from 
the parent active region to the footpoint of STEREO-B (top panel)
and Earth (bottom panel). EUV wave decayed before reaching Earth's footpoint. 
(A color version of this figure is available in the online journal).
\label{fig9}}
\end{figure}

\begin{figure}
\epsscale{.80}
\plotone{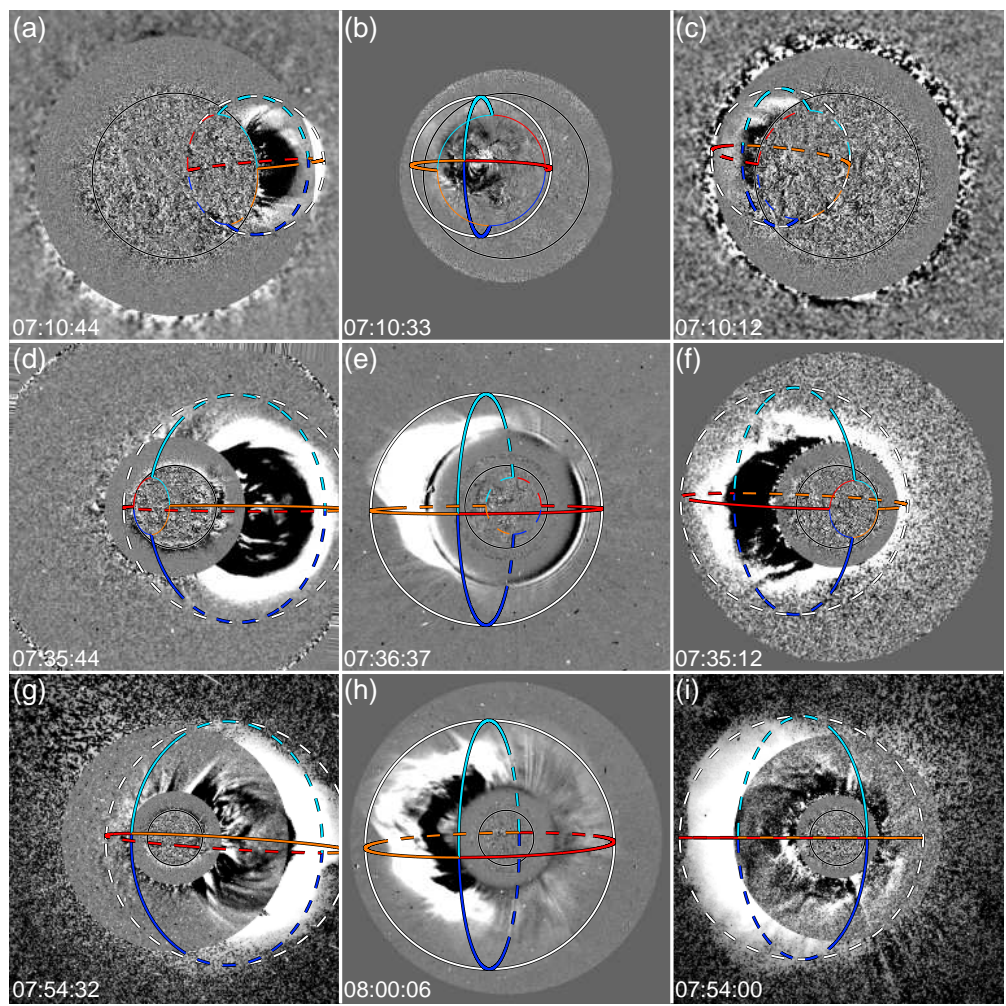}
\caption{Selected time series observations of the CME. Left, middle and right columns
show composite images observed from STEREO-B, SDO and SOHO, and STEREO-A, respectively.
The solar center is located at the center of each panel and the solar rotational axis is the north of each image.
The thin black circle in each panel refers to the solar disk.
Images at the center are running difference images from STEREO/SECCHI/EUVI 195 {\AA}  in the left and right columns
and running ratio images from SDO/AIA 193 {\AA} in the middle column.
White light observations are running difference images from STEREO/SECCHI/COR-1 and STEREO/SECCHI/COR-2 in the left and right columns
and SOHO/LASCO/C2 in the middle column.
The representations of the reconstructed 3D shock front are indicated by  the
red, orange, blue and cyan  lines whereas the white circle shows the maximum extension of the shock front.
Dashed lines are used when the 3D structure is located in the other side of the plane of the image. 
(A color version of this figure is available in the online journal).
\label{fig10}}
\end{figure}

\begin{figure}
\epsscale{.80}
\plotone{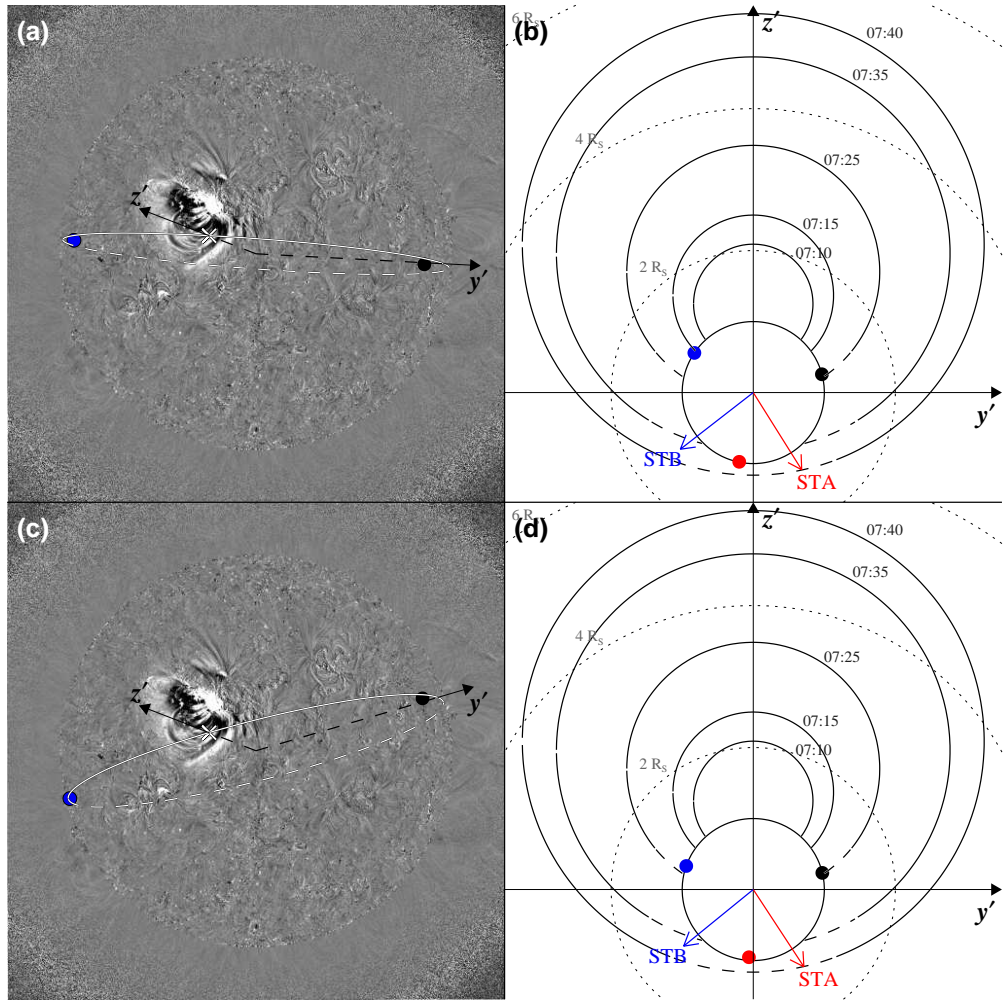}
\caption{Temporal evolution of the CME-driven shock determined with the ellipsoid and GCS models.
(Left) Schematic representation of the plane used to project the  outermost front shock.
The plane y'-z' is defined from three points on the solar surface: the center of the ellipsoid projected radially on the solar surface (white cross) 
and the magnetic footpoints of STEREO-B (blue dot) and L1 (black dot) estimated using the Parker Spiral method (top) and the MAS method (bottom).
(Right) Temporal evolution projected in the plane y'-z' of the outermost front (black traces) determined with the ellipsoid--GCS model.
Black dashed lines in the shock front trace are used below 1.5 R$_{\odot}$ when the EUV wave front cannot be tracked.
(A color version of this figure is available in the online journal).
\label{fig11}}
\end{figure}

\begin{figure}
\epsscale{.60}
\plotone{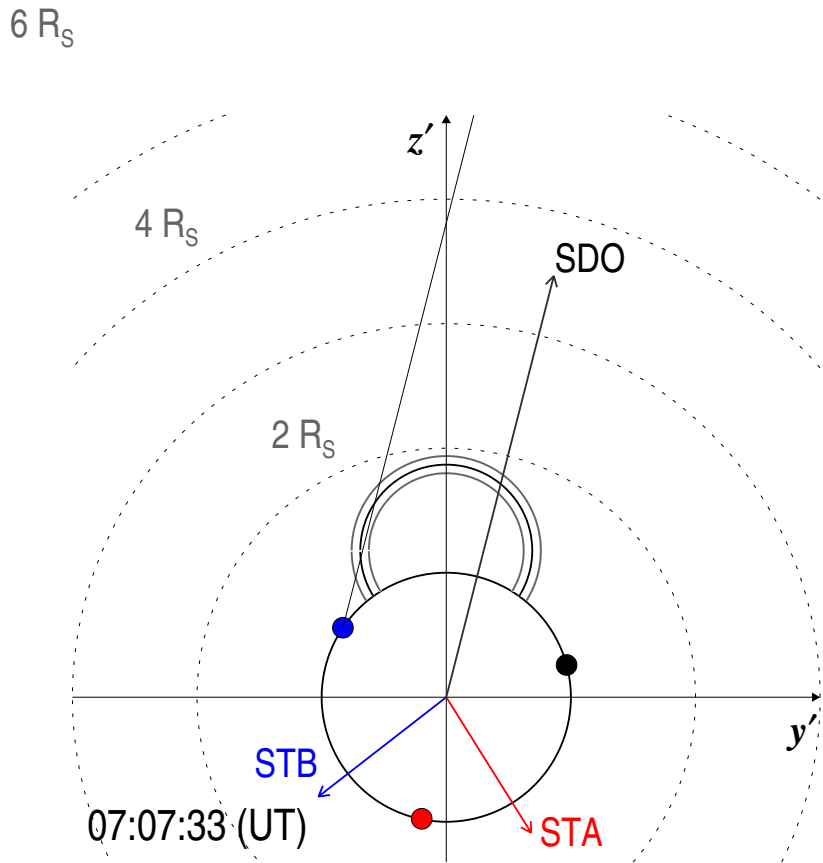}
\caption{Schematic configuration of the expanding coronal shock (black trace) at 07:07 UT.
The gray contours around the shock trace indicate the 10$\%$  uncertainty in the geometry of the modeled shock front.
The black, red and blue dots on the Sun surface (black circle at 1 R$_{S}$) indicate the footpoints of the magnetic field lines connecting to Earth, STEREO-A and STEREO-B, respectively.
The black, red and blue arrows point in the directions of SDO, STEREO-A (STA) and STEREO-B (STB), respectively.
The solid straight line indicate the line of sight from SDO passing through the STEREO-B footpoint.
Whereas the CME-driven shock has not reached the STEREO-B footpoint, a near-Earth observer will see its projection reaching this footpoint.
(A color version of this figure is available in the online journal).
\label{fig12}}
\end{figure}

\clearpage

\begin{table}
\begin{center}
\caption{Spacecraft and Magnetic Footpoint Locations.\label{tbl-1}}
\begin{tabular}{lrrrrccrcrrcrr}
\tableline\tableline
Observer & \multicolumn{4}{c}{S/C Location} & \multicolumn{1}{c}{$V_{sw}$} & \multicolumn{2}{c}{Parker spiral} & &  \multicolumn{2}{c}{MAS} & & \multicolumn{2}{c}{PFSS} \\
         \cline{2-5} \cline{7-8} \cline{10-11} \cline{13-14} 
                     & $R$ [AU] &  $Long$ & $CL$ & $Lat$  & [km s$^{-1}$] & $Long$ & $Lat$  & & $Long$ & $Lat$  & & $Long$ & $Lat$  \\
  \cline{2-5} \cline{7-8} \cline{10-11} \cline{13-14}
 \multicolumn{1}{c}{(1)} & \multicolumn{1}{c}{(2)} & \multicolumn{1}{c}{(3)}  & \multicolumn{1}{c}{(4)} & \multicolumn{1}{c}{(5)} & \multicolumn{1}{c}{(6)}  & \multicolumn{1}{c}{(7)} &  \multicolumn{1}{c}{(8)} &  & \multicolumn{1}{c}{(9)} & \multicolumn{1}{c}{(10)} & & \multicolumn{1}{c}{(11)} &  \multicolumn{1}{c}{(12)} \\
 \tableline
 STEREO-B &   1.02 &      344$^{\circ}$  & 304$^{\circ}$ & +2$^{\circ}$ & 339 & 55$^{\circ}$ &   +2$^{\circ}$ & & 41$^{\circ}$ & -15$^{\circ}$ & &  35$^{\circ}$ & -16$^{\circ}$ \\   
 Earth &   1.00  &      125$^{\circ}$  & 85$^{\circ}$ & -6$^{\circ}$ & 391 & 186$^{\circ}$ &  -6$^{\circ}$ & & 187$^{\circ}$ & +13$^{\circ}$ & & 187$^{\circ}$ & +13$^{\circ}$ \\
  STEREO-A &   0.96 &      259$^{\circ}$ & 219$^{\circ}$  & +7$^{\circ}$ & 529 & 302$^{\circ}$ &   +7$^{\circ}$ & & 299$^{\circ}$ & +16$^{\circ}$ & & 295$^{\circ}$ & +17$^{\circ}$ \\     
\tableline
\end{tabular}
\end{center}
\end{table}

\clearpage

\begin{table}
\begin{center}
\caption{Shock Height at the Particle Release Times.\label{tbl-2}}
\begin{tabular}{lcrrr}
\tableline\tableline
\multicolumn{1}{l}{Particle Species/Spacecraft/Instrument} & & \multicolumn{1}{c}{Estimated} & \multicolumn{1}{c}{Shock Height} & $\theta_{nr}$ \\
  &   & \multicolumn{1}{c}{Release} & \multicolumn{1}{c}{above Sun} & \\
 & & \multicolumn{1}{c}{Time [UT]$^{a}$} & \multicolumn{1}{c}{Surface$^{b}$} & \\
\multicolumn{1}{c}{(1)} & & \multicolumn{1}{c}{(2)} & \multicolumn{1}{c}{(3)}  & \multicolumn{1}{c}{(4)}  \\
\tableline
Protons/STEREO-B/HET &  &  07:10$\pm$4 min (VDA) &      $^{<}_{\sim}$0.13  R$_{\odot}$& 74$\pm$04$^{\circ}$  \\   
375-425 keV Electrons/STEREO-B/SEPT &  &  07:16$\pm$2 min (TSA) &  $^{<}_{\sim}$0.62 R$_{\odot}$&  69$\pm$13$^{\circ}$   \\
0.7-1.4 MeV Electrons/STEREO-B/HET &  &  07:17$\pm$2 min (TSA) &   $^{<}_{\sim}$0.72 R$_{\odot}$  & 69$\pm$14$^{\circ}$  \\
60-100 MeV Protons/STEREO-B/HET  &    & 07:26$\pm$3 min (TSA) & 1.20$\pm$0.80 R$_{\odot}$  & 53$\pm$16$^{\circ}$  \\   
\tableline
Protons/SOHO/ERNE &  &  07:58$\pm$9 min (VDA) &      3.68$\pm$1.27 R$_{\odot}$ & 33$\pm$06$^{\circ}$  \\   
 Near-relativistic Electrons/WIND/3DP &  &  07:45$\pm$5 min (VDA) & 2.57$\pm$0.90 R$_{\odot}$ &  36$\pm$07$^{\circ}$   \\
175-315 keV Electrons/ACE/EPAM/DE & &   07:47$\pm$2 min  (TSA) & 2.73$\pm$0.65 R$_{\odot}$ &   35$\pm$05$^{\circ}$   \\
0.25-0.70 MeV Electrons/SOHO/EPHIN &  &  07:35$\pm$2 min (TSA) & 1.52$\pm$0.63 R$_{\odot}$     & 45$\pm$08$^{\circ}$  \\
\tableline
\end{tabular}
\tablenotetext{a}{$\,$Light travel time already added. The method used to estimate the particle release time 
is indicated by VDA  for velocity dispersion analysis and TSA for time-shifted along the nominal Parker spiral length.}
\tablenotetext{b}{Shock height estimated in the radial direction above the footpoint.}
\end{center}
\end{table}


\clearpage






\end{document}